\documentclass[10pt,final,journal]{IEEEtran}
\usepackage[switch]{lineno} 
\usepackage{tabularx} 
\usepackage{booktabs}
\usepackage{multirow}
\usepackage{makecell}

\usepackage{stmaryrd}
\usepackage{graphicx}
\usepackage{amsthm} 
\usepackage{amsmath} 
\usepackage{bm}
\usepackage{graphicx}
\usepackage{epstopdf}
\usepackage{color}
\usepackage{float}
\usepackage[colorlinks=false,linkcolor=black,urlcolor=black,bookmarksopen=true, hidelinks]{hyperref}
\usepackage{bookmark}
\usepackage[flushleft]{threeparttable}
\usepackage{stackengine}
\usepackage{scalerel}
\usepackage{array}

\usepackage{xcolor,soul}

\usepackage{cite}

\ifCLASSINFOpdf
\else
\fi
\usepackage{amsfonts,amssymb}

\usepackage{amsmath}
\usepackage{algorithm}
\usepackage{algpseudocode}
\ifCLASSOPTIONcompsoc
 \usepackage[caption=false,font=normalsize,labelfont=sf,textfont=sf]{subfig}
\else
 \usepackage[caption=false,font=footnotesize]{subfig}
\fi
\newtheorem{theorem}{Theorem}

\begin{document}
%

\title{Optimal Two-way TOA Localization and Synchronization for Moving User Devices with Clock Drift}

\author{Sihao~Zhao, 
		Xiao-Ping~Zhang, \textit{Fellow, IEEE,}
        Xiaowei~Cui,
        and~Mingquan~Lu
\thanks{This work was supported in part by the Natural Sciences and Engineering Research Council of Canada (NSERC), Grant No. RGPIN-2020-04661. \textit{(Corresponding author: Xiao-Ping Zhang)}}
\thanks{Copyright (c) 2015 IEEE. Personal use of this material is permitted. However, permission to use this material for any other purposes must be obtained from the IEEE by sending a request to pubs-permissions@ieee.org.}
\thanks{S. Zhao, X.-P. Zhang are with the Department of Electrical, Computer and Biomedical Engineering, Ryerson University, Toronto, ON M5B 2K3, Canada (e-mail: sihao.zhao@ryerson.ca; xzhang@ryerson.ca).}
\thanks{X. Cui is with the Department of Electronic Engineering,
	Tsinghua University, Beijing 100084, China (e-mail: cxw2005@tsinghua.edu.cn).}
\thanks{M. Lu is with the Department of Electronic Engineering,
	Beijing National Research Center for Information Science and Technology, Tsinghua University, Beijing 100084, China. (e-mail: lumq@tsinghua.edu.cn).}
}

\markboth{}%
{Shell \MakeLowercase{\textit{et al.}}: Bare Demo of IEEEtran.cls for IEEE Journals}
%




\maketitle

\begin{abstract}
In two-way time-of-arrival (TOA) systems, a user device (UD) obtains its position and timing information by round-trip communications to a number of anchor nodes (ANs) at known locations. Compared with the one-way TOA technique, the two-way TOA scheme is easy to implement and has higher localization and synchronization accuracy. Existing two-way TOA methods assume a stationary UD. This will cause uncompensated position and timing errors. In this article, we propose an optimal maximum likelihood (ML) based two-way TOA localization and synchronization method, namely TWLAS. Different from the existing methods, it takes the UD mobility into account to compensate the error caused by the UD motion. We analyze its estimation error and derive the Cram\'er-Rao lower bound (CRLB). We show that the conventional two-way TOA method is a special case of the TWLAS when the UD is stationary, and the TWLAS has high estimation accuracy than the conventional one-way TOA method. We also derive the estimation error in the case of deviated UD velocity information. Numerical result demonstrates that the estimation accuracy of the new TWLAS for a moving UD reaches CRLB, better than that of the conventional one-way TOA method, and the estimation error caused by the deviated UD velocity information is consistent with the theoretical analysis.

\end{abstract}



%
\IEEEpeerreviewmaketitle

\section{Introduction}\label{Introduction}
%
%
%
%
\IEEEPARstart{I}{N} order to determine the position and clock offset of a user device (UD) in a wireless localization system, measurements such as time-of-arrival (TOA), angle-of-arrival (AOA), received signal strength (RSS) and a combination of them \cite{kuutti2018survey,wang2012novel,zhao2020closed,shi2019blas,shao2014efficient,luo2019novel,an2020distributed,coluccia2019hybrid,katwe2020nlos,tomic2019linear,zhao2021parn} have to be obtained with respect to anchor nodes (ANs) at known coordinates. Among these measurements, TOA has high accuracy and is used in a number of real-world applications such as global navigation satellite systems (GNSSs), ultra-wide band (UWB) indoor positioning, and smart vehicle autonomous navigation \cite{zafari2019survey,lu2019overview,zhao2014kalman,conti2019soft,wu2019coordinate}.

The TOA-based localization and synchronization techniques are usually categorized into one-way and two-way TOA schemes \cite{liu2007survey,guvenc2009survey}. One-way TOA is referred to as obtaining the TOA measurements by recording the timestamps of the one-way signal transmission and reception based on the local clock sources of the AN and UD \cite{yan2013review,shi2020sequential}. In two-way TOA, two one-way range measurements are obtained from timestamps of each round-trip communication between the AN and the UD. Compared with the one-way TOA scheme, two-way TOA requires more communication times but has better localization and synchronization accuracy due to more available TOA measurements \cite{zafari2019survey}. 

In a two-way TOA system, a UD communicates with a number of ANs in a round-trip manner to obtain sufficient amount of TOA measurements for UD localization and synchronization \cite{bialer2016two}. This communication protocol is straightforward and easy to be implemented. Thus, it has been extensively studied and a variety of two-way TOA localization and synchronization methods are presented in literature \cite{bialer2016two,gholami2016tw,lazzari2017numerical,zheng2010joint,vaghefi2015cooperative,zou2017joint,nevat2016location,tomic2018exact,yuan2016cooperative,yin2018gnss,gao2016robust}.

These previous studies all assume that the target or UD to be located is stationary. This assumption can hold in applications such as wireless sensor networks where all sensors are placed at fixed positions. However, for other dynamic applications such as drone navigation, smart vehicle control and personnel tracking, ignoring the UD motion will result in extra position and timing errors, which seriously degrade the performance of localization and synchronization.

In this article, we develop a new optimal two-way TOA localization and synchronization method for a moving UD with clock drift, namely TWLAS, which compensates the localization and synchronization error caused by the UD motion and has higher accuracy than the conventional two-way TOA methods. Unlike existing two-way TOA methods, which do not take the UD motion into account, we formulate the localization and synchronization problem for a moving UD by modeling the UD motion with a constant velocity during a short period. We present an iterative algorithm for the TWLAS method. We derive the CRLB of the new TWLAS in the two cases with and without known UD velocity. We show that the TWLAS outperforms the conventional two-way TOA and one-way TOA methods in localization and synchronization accuracy. Numerical simulations show that the estimation accuracy of the TWLAS reaches CRLB. For a moving UD, the new TWLAS method compensates the localization and synchronization error caused by the UD motion and significantly outperforms the conventional two-way TOA method. All the numerical results are consistent with theoretical analysis.

The rest of the article is organized as follows. In Section II, the TOA measurements and the UD motion and clock are modeled, and the two-way TOA localization problem is formulated. In Section III, the optimal localization method, namely TWLAS, as well as its iterative algorithm are proposed. The estimation error of the proposed TWLAS method is analyzed in Section IV. Numerical simulations are conducted to evaluate the performance of the TWLAS method in Section V. Section VI concludes this article.
 
Main notations are summarized in Table \ref{table_notation}.

\begin{table}[!t]
	\caption{Notation List}
	\label{table_notation}
	\centering
	\begin{tabular}{l p{5.5cm}}
		\toprule
		lowercase $x$&  scalar\\
		bold lowercase $\boldsymbol{x}$ & vector\\
		bold uppercase $\bm{X}$ & matrix\\
		$\Vert \boldsymbol{x} \Vert$ & Euclidean norm of a vector\\
		$\Vert \boldsymbol{x}\Vert _{\bm{W}}^2$ & square of Mahalanobis norm, i.e., $\boldsymbol{x}^T\bm{W}\boldsymbol{x}$\\
		$i$, $j$ & indices of variables\\
		$[\boldsymbol{x}]_{i}$ &the $i$-th element of a vector\\
		$\mathrm{tr}(\bm{X})$ & trace of a matrix\\
		$[\bm{X}]_{i,:}$, $[\bm{X}]_{:,j}$ &the $i$-th row and the $j$-th column of a matrix, respectively\\
		$[\bm{X}]_{i,j}$ &entry at the $i$-th row and the $j$-th column of a matrix\\
		$[\bm{X}]_{i:m,j:n}$ &sub-matrix from the $i$-th to the $m$-th row and from the $j$-th to the $n$-th column of a matrix\\
		$\mathbb{E}[\cdot]$ & expectation operator \\
		$\mathrm{diag}(\cdot)$ & diagonal matrix with the elements inside\\
		$M$ & number of ANs\\
		$N$ & dimension of all the position and velocity vectors, i.e., $N=2$ in 2D case and $N=3$ in 3D case\\
		$\bm{I}_M$ & $M\times M$ identity matrix\\
		$\bm{O}_{M\times N}$ & $M\times N$ zero matrix\\
		$\boldsymbol{0}_{M}$, $\boldsymbol{1}_{M}$& $M$-element vectors with all-zero and all-one elements, respectively\\
		$\boldsymbol{p}_{i}$ & position vector of AN \#$i$\\
		$\boldsymbol{p}$ &  unknown position vector of UD\\
		$\boldsymbol{v}$ &  velocity vector of UD\\
		$b$, $\omega$ &  unknown UD clock offset and clock drift \\
		$\boldsymbol{e}$, $\boldsymbol{l}$ & unit line-of-sight (LOS) vector from the UD to the AN at the UD transmission and reception time, respectively\\
		$\delta t_i$ & interval between UD signal transmission and reception from AN \#$i$ \\
		$c$ & propagation speed of the signal\\
		$\rho_{i}$ & request-TOA measurement at AN \#$i$ upon AN reception of the request signal from the UD\\
		$\tau_{i}$ & response-TOA measurement at the UD upon reception of the response signal from AN \#$i$\\
		$\boldsymbol{\theta}$ &  parameter vector\\
		$\varepsilon$, $\sigma^2$ &  Gaussian random error and variance\\
		$\mathcal{F}$ &  Fisher information matrix (FIM)\\
		$\bm{W}$ & weighting matrix\\
		$\bm{G}$ & design matrix\\
		$\mu$& estimation bias\\
		$\bm{Q}$ & estimation error variance matrix\\
		\bottomrule
	\end{tabular}
\end{table}

\section{Problem Formulation} \label{problem}
\subsection{Two-way TOA System Model}
\begin{figure}
	\centering
	\includegraphics[width=1\linewidth]{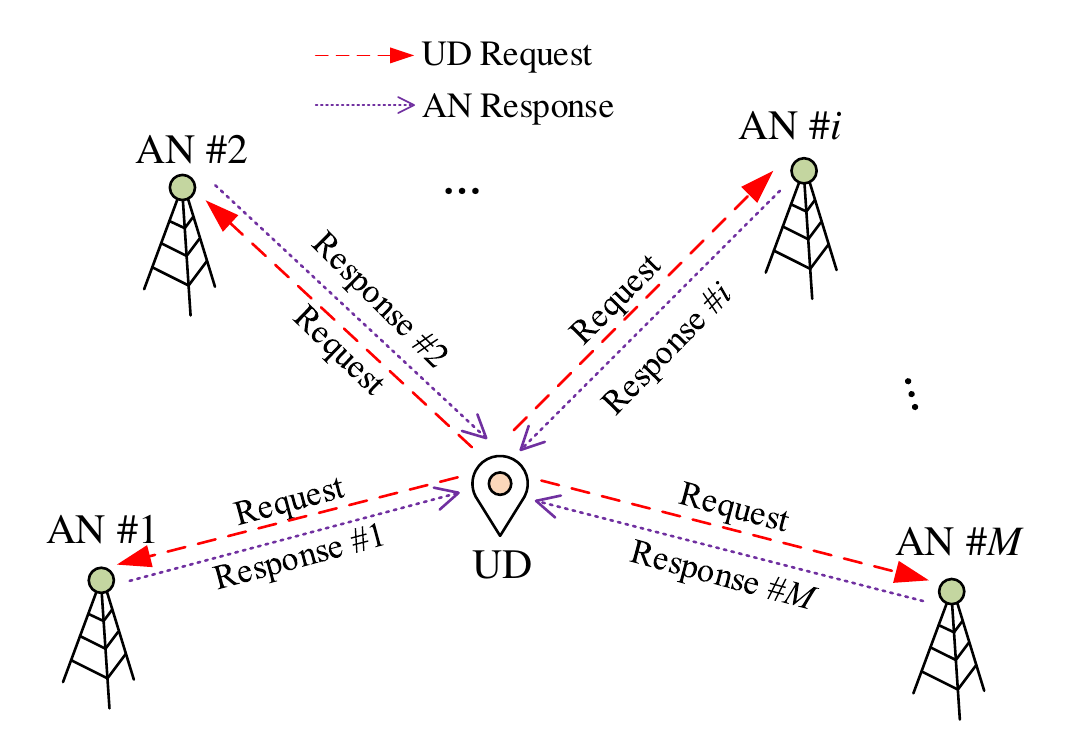}
	\caption{Two-way TOA localization and synchronization system. The moving UD transmits the request signal, and all ANs receive. $M$ request-TOA measurements are formed at all ANs. Then the ANs transmit response signal sequentially to avoid collision. $M$ sequential response-TOA measurements are formed at the UD.
	}
	\label{fig:systemfig}
\end{figure}

In the two-way TOA system as shown in Fig. \ref{fig:systemfig}, there are $M$ ANs placed at known positions. The coordinate of AN \#$i$ is denoted by $\boldsymbol{p}_i$, $i=1,\cdots,M$. The ANs are all synchronous, i.e., the clock offset and drift between any AN pair are known. This can be achieved by conducting multiple communications between ANs \cite{shi2019blas}. The UD position, denoted by $\boldsymbol{p}$, and clock offset, denoted by $b$ are unknowns to be determined. Both $\boldsymbol{p}_i$ and $\boldsymbol{p}$ are of $N$ dimension ($N=2$ for 2D cases and $N=3$ for 3D cases), i.e., $\boldsymbol{p}_{i} \text{, } \boldsymbol{p} \in \mathbb{R}^{N}$.

As shown in Fig. \ref{fig:systemfig}, during a localization and synchronization period, the UD transmits the request signal and all ANs receive it. Thus, $M$ TOA measurements are formed at the AN ends, namely request-TOA. AN \#$i$ processes the received signal and then transmits the response signal and the UD receives it to form TOA measurements, namely response-TOA. Once all the $M$ ANs finish transmission, $M$ response-TOA measurements are formed at the UD end. The communication protocol between the UD and ANs can be alternated, e.g., the UD can transmit and then receive signal to and from each AN at a time, but such alternation does not affect how the method proposed in Section \ref{locmethod} works.

\subsection{UD Motion and Clock Model}\label{clockandmovemodel}
The clock offset and drift of the UD with respect to the synchronous ANs are denoted by $b$ and $\omega$, respectively. Following the clock model in \cite{zou2017joint,hasan2018gnss}, we model the UD drift as a constant during a short period and the clock offset as the integration of the clock drift. It is expressed by
\begin{equation} \label{eq:clockbomega}
b(t_2) =b(t_1)+\omega(t_1)\cdot (t_2-t_1)\text{,}
\end{equation}
where $t_1$ and $t_2$ are two time instants close enough to ensure $\omega$ is constant during the interval.

We model the UD motion with a constant velocity. Specifically, we assume that the UD velocity, denoted by $\boldsymbol{v}$, remains stable during a short time period. Then, the UD motion is modeled as
\begin{equation} \label{eq:posvel}
\boldsymbol{p}(t_2) = \boldsymbol{p}(t_1) + \boldsymbol{v}(t_1) \cdot (t_2-t_1) \text{.}
\end{equation}

\subsection{Two-way TOA Localization and Synchronization}\label{lasproblem}
When the UD transmits the request signal and the ANs receive. The request-TOA measurement at AN \#$i$ ($i=1,\cdots,M$), denoted by $\rho_i$, equals the difference of the true signal propagation time and the clock offset plus measurement noise. Therefore, we have
\begin{align} \label{eq:rhoANi}
	\rho_i = t_{RX}^{(i)}- t_{TX}=
	\frac{\left\Vert\boldsymbol{p}_i-\boldsymbol{p}\right\Vert}{c} -b+ \varepsilon_{i} \text{, } i=1,\cdots,M \text{,}
\end{align}
where $t_{RX}^{(i)}$ is the local reception time at AN \#$i$, $t_{TX}$ is the UD local transmission time of the request signal, $\boldsymbol{p}$ and $b$ are the UD position and clock offset at $t_{TX}$, respectively, $t_{TX}$ is the UD transmission time of the request signal, $c$ is the signal propagation speed, and $\varepsilon_{i}$ is the measurement noise for AN \#$i$, following independent zero-mean Gaussian distribution with a variance of $\sigma_{i}^2$, i.e., $\varepsilon_{i} \sim \mathcal{N}(0,\sigma_{i}^2)$. The Gaussian distribution for measurement noise is widely adopted in literature \cite{zheng2010joint,vaghefi2015cooperative,zou2017joint,gao2016robust}. However, in practice, the measurements may deviate from this distribution due to interference such as impulse noises, which leads to large errors in TOA measurements. Preprocessing measures can be taken to detect and remove these erroneous measurements to ensure the correct localization and synchronization result \cite{enosh2014outlier,van2016optimizing,xiao2013robust}.

After receiving the request signal, all ANs transmit the response signal in a sequential manner. The UD receives the response signal from AN \#$i$ and forms a response-TOA measurement, denoted by $\tau_i$. The interval from the UD transmission to the reception of the response signal from AN \#$i$ is denoted by $\delta t_i$. We use the UD states including position, velocity, clock offset and clock drift at the transmission instant to express $\tau_i$ as
\begin{align} \label{eq:tauANi}
	\tau_i =t_{RX}-t_{TX}^{(i)}= \frac{\left\Vert\boldsymbol{p}_i-\boldsymbol{p}-\boldsymbol{v}\cdot\delta t_i\right\Vert}{c}+ &b+\omega\cdot\delta t_i + \varepsilon, \nonumber\\
	&i=1,\cdots,M \text{,}
\end{align}
where $t_{RX}$ is the local reception time of the response signal at the UD, $t_{TX}^{(i)}$ is the local transmission time at AN \#$i$, $\boldsymbol{p}$, $b$ and $\omega$ are all at the instant of $t_{TX}$, and $\varepsilon$ is the measurement noise for the UD, following a zero-mean Gaussian distribution with a variance of $\sigma^2$, i.e., $\varepsilon \sim \mathcal{N}(0,\sigma^2)$. 

Based on the measurements given by (\ref{eq:rhoANi}) and (\ref{eq:tauANi}), the problem of localization and synchronization for the UD is to estimate the position $\boldsymbol{p}$ and the clock offset $b$ at the instant $t_{TX}$. The estimation method will be proposed in Section \ref{locmethod}.

\section{Optimal Two-way TOA Localization and Synchronization} \label{locmethod}
In this section, we will develop a ML method, namely TWLAS, to achieve localization and synchronization for a moving UD. The iterative algorithm of the TWLAS will be presented as well.
\subsection{ML Estimator for Localization and Synchronization} \label{estimator}
The unknown parameters we are interested in for localization and synchronization are the UD position $\boldsymbol{p}$ and its clock offset $b$ at the instant $t_{TX}$. However, by observing (\ref{eq:tauANi}), we also need to handle the UD velocity $\boldsymbol{v}$ and clock drift $\omega$. In practice, the UD velocity can be obtained if the UD is stationary or a motion sensor such as an inertial measurement unit is equipped. Therefore, we consider two cases, one is with known UD velocity and the other is without. For the former case, the unknown parameters to be estimated include $\boldsymbol{p}$, $b$, and $\omega$, while for the latter case, the unknown parameters are $\boldsymbol{p}$, $b$, $\boldsymbol{v}$ and $\omega$. Correspondingly, we design two modes for the TWLAS to deal with the two cases. The unknown parameter vector is
\begin{align} \label{eq:thetadef}
\boldsymbol{\theta}=\left\{
\begin{matrix}
\left[\boldsymbol{p}^T,b,\omega\right]^T, & \text{for Mode 1,}\\
\left[\boldsymbol{p}^T,b,\omega,\boldsymbol{v}^T\right]^T, &\text{for Mode 2.}
\end{matrix}
\right.
\end{align}

We note that the unknowns to be estimated in Mode 1 are the same as the conventional two-way TOA method, such as presented in \cite{zheng2010joint,vaghefi2015cooperative,zou2017joint}. Therefore, the conventional two-way TOA method that estimates $\boldsymbol{p}$, $b$, and $\omega$, ignoring the UD motion, is a special case of Mode 1 when the UD is stationary. However, without employing the UD velocity, the conventional method will produce uncompensated error for a moving UD as shown in Section \ref{perfromancedV}.

The two-way TOA measurements are written in the collective form as
$$
\boldsymbol{\rho}=
\left[\rho_1,\cdots,\rho_M,\tau_1,\cdots,\tau_M\right]^T \text{.} 
$$

The relation between the unknown parameters and the measurements is
\begin{equation} \label{eq:rhoandtheta}
\boldsymbol{\rho} = h(\boldsymbol{\theta}) + \boldsymbol{\varepsilon} \text{,}
\end{equation}
where based on (\ref{eq:rhoANi}) and (\ref{eq:tauANi}), the $i$-th row of the function $h(\boldsymbol{\theta})$ is
\begin{align} \label{eq:funtheta}
&\left[h(\boldsymbol{\theta})\right]_{i} = \nonumber\\
&\left\{
\begin{matrix}
\frac{\left\Vert\boldsymbol{p}_i-\boldsymbol{p}\right\Vert}{c}-b, & i=1,\cdots,M, \\
\frac{\left\Vert\boldsymbol{p}_{i-M}-\boldsymbol{p}-\boldsymbol{v}\cdot\delta t_{i-M}\right\Vert}{c}+b+\omega \cdot \delta t_{i-M},&i=M+1,\cdots,2M,
\end{matrix} 
\right.
\end{align}
and $\boldsymbol{\varepsilon}=\left[\varepsilon_1,\cdots,\varepsilon_{M},\varepsilon\boldsymbol{1}_M^T\right]^T$ with $\boldsymbol{1}_M$ being an all-one $M$-vector.

According to the measurement model presented in the previous section, all the error terms are independently Gaussian distributed. The ML estimation of $\boldsymbol{\theta}$ is written as a weighted least squares (WLS) minimizer as
\begin{equation} \label{eq:MLminimizer}
\hat{\boldsymbol{\theta}}=\text{arg}\min\limits_{{\boldsymbol{\theta}}} \left\Vert\boldsymbol{\rho} - \mathit{h}({\boldsymbol{\theta}})\right\Vert_{\bm{W}}^2
\text{,}
\end{equation}
where $\hat{\boldsymbol{\theta}}$ is the estimator, and $\bm{W}$ is a diagonal positive-definite weighting matrix given by
\begin{equation} \label{eq:matW}
\bm{W}=
\left[
	\begin{matrix}
	\bm{W}_{\rho} & \bm{O}_{M\times M}\\
	\bm{O}_{M\times M}& \bm{W}_{\tau}
	\end{matrix}
\right]\text{,}
\end{equation}
in which $\bm{O}_{M\times M}$ is a $M\times M$ square matrix with all entries being zero, and
\begin{equation} \label{eq:matWrho}
\bm{W}_{\rho}=\mathrm{diag}\left(\frac{1}{\sigma_1^2},\cdots,\frac{1}{\sigma_M^2}\right) \text{,}
\end{equation}
\begin{equation} \label{eq:matWtau}
	\bm{W}_{\tau}= \frac{1}{\sigma^2}\bm{I}_M\text{,}
\end{equation}
with $\bm{I}_M$ being an identity matrix.

\subsection{Iterative Localization and Synchronization Algorithm}
In order to solve the minimization problem given by (\ref{eq:MLminimizer}), we develop an iterative algorithm for the proposed TWLAS method, following the Gauss-Newton algorithm in \cite{kaplan2005understanding,huang2015dilution,zhao2017priori}. We conduct Taylor series expansion on (\ref{eq:rhoandtheta}) at the estimate point of
$$
\check{\boldsymbol{\theta}}=\left\{
\begin{matrix}
\left[\check{\boldsymbol{p}}^T,\check{b},\check{\omega}\right]^T & \text{for Mode 1,}\\
\left[\check{\boldsymbol{p}}^T,\check{b},\check{\omega},\check{\boldsymbol{v}}^T\right]^T &\text{for Mode 2,}
\end{matrix}
\right.
$$
where $\check{\boldsymbol{p}}$, $\check{b}$, $\check{\omega}$, and $\check{\boldsymbol{v}}$ are estimates for $\boldsymbol{p}$, $b$, ${\omega}$, and ${\boldsymbol{v}}$, respectively. We keep the first-order term and ignore the higher order terms, and then have
\begin{equation} \label{eq:GNtaylor}
	\boldsymbol{\rho} = \mathit{h}(\check{\boldsymbol{\theta}}) + \check{\bm{G}}\cdot \Delta\boldsymbol{\theta}+\boldsymbol\varepsilon \text{,}
\end{equation}
where $\check{\bm{G}}$ is the estimation of the design matrix
$$\bm{G}=\frac{\partial \mathit{h}(\boldsymbol{\theta})}{\partial \boldsymbol{\theta}} \text{,}$$
$
\check{\bm{G}}=\frac{\partial \mathit{h}(\boldsymbol{\theta})}{\partial \boldsymbol{\theta}}|_{\boldsymbol{\theta}=\check{\boldsymbol{\theta}}}
$,
and $\Delta\boldsymbol{\theta}$ is the error vector given by
$\Delta\boldsymbol{\theta} = \boldsymbol{\theta}-\check{\boldsymbol{\theta}} \text{.}
$

The design matrices for the two modes of the TWLAS, denoted by $\bm{G}_{\text{Mode 1}}$ and $\bm{G}_{\text{Mode 2}}$, respectively, are
\begin{align} \label{eq:matG}
\bm{G}=\left\{
\begin{matrix}
\bm{G}_{\text{Mode 1}}=\left[
\begin{matrix}
\bm{G}_0 & \boldsymbol{0}_M\\
\bm{G}_1 & [\delta t_1,\cdots,\delta t_M]^T
\end{matrix}
\right] \text{,} &\text{for Mode 1,}\\
\bm{G}_{\text{Mode 2}}=\left[
\begin{matrix}
	\bm{G}_0 & \bm{O}_{M\times(N+1)}\\
	\bm{G}_1 & \bm{G}_2
\end{matrix}
\right]
\text{,} &\text{for Mode 2,}
\end{matrix}
\right.
\end{align}
where
$$
\bm{G}_0=\left[
\begin{matrix}
	-\boldsymbol{e}_1^T & -1 \\
	\vdots & \vdots \\
	-\boldsymbol{e}_M^T & -1
\end{matrix}
\right] \text{,}\;
\bm{G}_1=\left[
\begin{matrix}
	-\boldsymbol{l}_1^T & 1 \\
	\vdots & \vdots \\
	-\boldsymbol{l}_M^T & 1
\end{matrix}
\right] \text{,}
$$
$$
\bm{G}_2=\left[
\begin{matrix}
	\delta t_1  & -\boldsymbol{l}_1^T\delta t_1\\
	\vdots & \vdots \\
	\delta t_M & -\boldsymbol{l}_M^T\delta t_M
\end{matrix}
\right]\text{,}
$$
with $\boldsymbol{e}$ representing the unit line-of-sight (LOS) vector from the UD to the AN at the time instant of UD transmission,
\begin{equation} \label{eq:rhoLOS}
\boldsymbol{e}_{i}=\frac{\boldsymbol{p}_i - {\boldsymbol{p}} }{\Vert \boldsymbol{p}_i - {\boldsymbol{p}}\Vert } , i=1,\cdots,M,
\end{equation}
and
$\boldsymbol{l}$ representing the unit LOS vector from the UD to AN \#$i$ at the UD reception time as
\begin{equation} \label{eq:tauLOS}
\boldsymbol{l}_i=\frac{\boldsymbol{p}_i - {\boldsymbol{p}}-{\boldsymbol{v}}\cdot \delta t_i}{\Vert \boldsymbol{p}_i - {\boldsymbol{p}}-{\boldsymbol{v}} \cdot\delta t_i\Vert }, i=1,\cdots,M\text{,}
\end{equation}
and $\boldsymbol{0}_M$ is an all-zero $M$-vector.

The residual vector is denoted by $\boldsymbol{r}$, 
\begin{equation}\label{eq:residual}
\boldsymbol{r} = \boldsymbol{\rho} - \mathit{h}(\check{\boldsymbol{\theta}})= \check{\bm{G}} \cdot \Delta\boldsymbol{\theta}+\boldsymbol\varepsilon \text{.}
\end{equation}

We denote the WLS estimate of the error vector $\Delta\boldsymbol{\theta}$ by $\Delta \check{\boldsymbol{\theta}}$, and have
\begin{equation} \label{eq:leastsquare}
\Delta\check{\boldsymbol{\theta}}=(\check{\bm{G}}^T\bm{W}\check{\bm{G}})^{-1}\check{\bm{G}}^T\bm{W}\boldsymbol{r} \text{.}
\end{equation}

The unknown parameter vector to be estimated is thereby updated iteratively by
\begin{equation} \label{eq:esttheta}
\check{\boldsymbol{\theta}} \leftarrow \check{\boldsymbol{\theta}} + \Delta \check{\boldsymbol{\theta}} \text{.}
\end{equation}

We then update the matrix $\check{\bm{G}}$ and the residual $\boldsymbol{r}$ using the estimated parameter from (\ref{eq:esttheta}) iteratively until convergence. The iterative procedure is given by Algorithm 1.

The noise variances in the weighting matrix $\bm{W}$ are treated as known in the proposed method, similar to what is done in \cite{zheng2010joint,vaghefi2015cooperative, gao2016robust}. However, in practice, we need to take some measures to obtain the noise variance. For example, we can collect the TOA measurement data before the devices are put in use and identify the noise variance by fitting the collected data. Then the identified value can be used in the real applications. The study in \cite{coluccia2019hybrid} provides an effective way to estimate the TOA measurement noise through a calibration step using the TOA messages between ANs.
	
Note that the proposed iterative algorithm requires a proper initial guess to guarantee convergence to the correct solution. In practice, some prior knowledge such as a rough estimate or the previous value of the UD position can be used as the initial guess. We will evaluate the dependence of the algorithm on the initial parameter value in the next section.


\begin{algorithm}
	\caption{Iterative TWLAS Algorithm}
	\begin{algorithmic}[1]
		\State Input: TOA measurements $\boldsymbol{\rho}$ and weighting matrix $\bm{W}$, ANs' positions $\boldsymbol{p}_i$, $i=1,\cdots,M$, UD velocity $\boldsymbol{v}$ (for Mode 1), initial parameter estimate $\check{\boldsymbol{\theta}}_{0}$, maximum iteration count $iter$, and convergence threshold $thr$
		\For {$k=1:iter$}
		\State Compute LOS vectors $\boldsymbol{e}_{i}$ and $\boldsymbol{l}_{i}$, $i=1,\cdots,M$, based on (\ref{eq:rhoLOS}) and (\ref{eq:tauLOS})
		\State Calculate residual $\boldsymbol{r}$ using (\ref{eq:residual})
		\State Form design matrix $\check{\bm{G}}$ based on (\ref{eq:matG})
		\State Calculate estimated error $\Delta \check{\boldsymbol{\theta}}$ using (\ref{eq:leastsquare})
		\State Update parameter estimate $\check{\boldsymbol{\theta}}_{k} = \check{\boldsymbol{\theta}}_{k-1} + \Delta \check{\boldsymbol{\theta}}$
		\If {$\left\Vert [\Delta \check{\boldsymbol{\theta}}]_{1:N} \right\Vert<thr$}
		\State Exit \textbf{for} loop
		\EndIf
		\EndFor
		\State Output: $\check{\boldsymbol{\theta}}_{k}$
	\end{algorithmic}
\end{algorithm}

\section{Estimation Error Analysis} \label{locanalysis}

%
%
%
%

\subsection{CRLB Analysis}\label{CRLBanalysis}
CRLB is the lower bound for the covariance of an unbiased estimator. It is calculated from the inverse of the Fisher information matrix (FIM) as given by
\begin{equation} \label{eq:CRLBFisher}
	\mathsf{CRLB}([\boldsymbol{\theta}]_i)=[\mathcal{F}^{-1}]_{i,i} \text{,}
\end{equation}
where $\mathcal{F}$ is the FIM, $[\cdot]_{i}$ represents the $i$-th element of a vector, and $[\cdot]_{i,j}$ represents the entry at the $i$-th row and the $j$-th column of a matrix.


The FIM of the TWLAS is written by 
\begin{equation} \label{eq:FisherExpectation}
\mathcal{F}=\left(\frac{\partial h(\boldsymbol{\theta})}{\partial \boldsymbol{\theta}}\right)^T\bm{W}\frac{\partial h(\boldsymbol{\theta})}{\partial \boldsymbol{\theta}}= \bm{G}^T\bm{W}\bm{G} \text{.}
\end{equation}

The parameters to be estimated for the two modes of the TWLAS, as defined by (\ref{eq:thetadef}) in Section \ref{estimator}, are different. We denote the FIMs of the two modes by $\mathcal{F}_{\text{Mode 1}}$ for Mode 1 and $\mathcal{F}_{\text{Mode 2}}$ for Mode 2, respectively. The terms that relate to localization and synchronization accuracy are the diagonal elements in the top-left $(N+1)\times(N+1)$ sub-matrix of the inverse of the FIMs, i.e.$\left[\mathcal{F}_{\text{Mode 1}}^{-1}\right]_{i,i}$ and $\left[\mathcal{F}_{\text{Mode 2}}^{-1}\right]_{i,i}$, $i=1,\cdots,N+1$. We have the following theorem.


\begin{theorem}\label{theorem0}
	The localization and synchronization accuracy of TWLAS Mode 1 is higher than that of Mode 2, i.e.,
\begin{equation} \label{eq:Fisher1and2}
	\left[\mathcal{F}_{\text{Mode 1}}^{-1}\right]_{i,i}<\left[\mathcal{F}_{\text{Mode 2}}^{-1}\right]_{i,i}, i=1,\cdots,N+1 \text{.}
\end{equation}
\end{theorem}
\textit{Proof.} See Appendix \ref{Appendix1}.

\textbf{Remark 1}: For TWLAS Mode 1, the UD motion, i.e., $\boldsymbol{v}$ is known and we only need to estimate the UD position, clock offset and drift. This helps Mode 1 to achieve better accuracy than Mode 2.

\subsection{Comparison with Conventional Two-way TOA Method} \label{compareold}
Different from the proposed TWLAS method, the conventional two-way TOA method presented by \cite{zheng2010joint,vaghefi2015cooperative,zou2017joint}, namely CTWLAS, ignores the UD movement and only estimates the UD position and clock parameters, i.e., $\boldsymbol{p}$, $b$, and $\omega$. For a moving UD, there will be estimation errors. We denote the estimation bias and the RMSE of the CTWLAS by $\boldsymbol{\mu}_C$ and $RMSE_C$, respectively. They are given by
\begin{equation}\label{eq:dPvsdVold}
	{\boldsymbol{\mu}}_C=(\bm{G}_{C}^T\bm{W}{\bm{G}_C})^{-1}\bm{G}_C^T\bm{W}{\boldsymbol{r}_C} \text{,}
\end{equation}
and
\begin{align} \label{eq:rmseold}
	RMSE_C =\sqrt{\Vert\boldsymbol{\mu}_C\Vert^2+\mathrm{tr}\left(\left(\bm{G}_C^T\bm{W}{\bm{G}_C}\right)^{-1}\right)} \text{,}
\end{align}
where $\bm{G}_C=\bm{G}_\text{Mode 1}$, and
\begin{align}\label{eq:resold}
{\boldsymbol{r}}_C
	=\left[
	\begin{matrix}
		\boldsymbol{0}_M\\
		\left\Vert \boldsymbol{p}_1 - \boldsymbol{p}  \right\Vert  -\left\Vert \boldsymbol{p}_1 - {\boldsymbol{p}} - {\boldsymbol{v}}  \delta t_1\right\Vert \\
		\vdots\\
		\left\Vert \boldsymbol{p}_M - \boldsymbol{p} \right\Vert  -\left\Vert \boldsymbol{p}_M - {\boldsymbol{p}} - {\boldsymbol{v}}  \delta t_M\right\Vert
	\end{matrix}
	\right]
	\text{.}
\end{align}

We can see that the term $\Vert\boldsymbol{\mu}_C\Vert^2$ is the extra error of the CTWLAS caused by the UD velocity. With increasing $\boldsymbol{v}$ and $\delta t_i$, the estimation error will increase. Therefore, for a moving UD, the localization and synchronization error of the CTWLAS is larger than that of Mode 1 of the proposed TWLAS method.

\subsection{Comparison with Conventional One-way TOA Method}
In order to obtain some insights on the estimation performance of the TWLAS Mode 2, we compare it with the commonly adopted conventional one-way TOA localization and synchronization method \cite{foy1976position,kaplan2005understanding}, namely OWLAS. The OWLAS only uses half of the measurements compared with the TWLAS. The unknown parameters to be estimated by the OWLAS include position $\boldsymbol{p}$ and clock offset $b$ only, also less than the TWLAS. Thus, it is not straightforward to obtain an intuition on which method has better estimation accuracy. We note that $\bm{G}_0$ equals to the design matrix for the conventional OWLAS method. We denote the FIM for the OWLAS by $\mathcal{F}_\text{OWLAS}$, and have
\begin{equation} \label{eq:Fisheroneway}
\mathcal{F}_{\text{OWLAS}}= \bm{G}_0^T\bm{W}\bm{G}_0\text{,}
\end{equation}

\begin{theorem}\label{theorem1}
	The localization and synchronization accuracy of TWLAS Mode 2 is higher than or equal to that of the conventional one-way TOA localization (OWLAS), i.e.,
	\begin{equation} \label{eq:Fisher2andOW}
		\left[\mathcal{F}_{\text{Mode 2}}^{-1}\right]_{i,i}\leq \left[\mathcal{F}_{\text{OWLAS}}^{-1}\right]_{i,i},i=1,\cdots\,N+1 \text{.}
	\end{equation}
\end{theorem}
\textit{Proof.} See Appendix \ref{Appendix2}.

\textbf{Remark 2}: In the case that the UD receives the response signals from all the ANs simultaneously, the equality in (\ref{eq:Fisher2andOW}) holds, as shown in Appendix \ref{Appendix2}, i.e., the estimation accuracy of the TWLAS Mode 2 is the same as that of the conventional OWLAS. However, in practice, for consumer level devices such as IoT systems, we should design a proper communication protocol to avoid possible air collision of such concurrent signals as well as reduce the power consumption and complexity in signal processing of the UD.


According to Theorems \ref{theorem0} and \ref{theorem1}, we have shown that the proposed TWLAS method has better estimation accuracy than that of the conventional OWLAS method.

\subsection{Estimation Error of Mode 1 caused by Deviated UD Velocity} \label{deviatedV}
In real-world applications, the obtained UD velocity information may not be accurate. For Mode 2 of the TWLAS, the position and clock offset of the UD are estimated regardless of the UD velocity. Thus, the inaccurate UD velocity information does not influence the estimation error of Mode 2. However, this inaccuracy will cause localization and synchronization error to the TWLAS Mode 1 as will be analyzed in this sub-section.

We denote the obtained UD velocity by $\tilde{\boldsymbol{v}}$. The deviation from the true velocity $\boldsymbol{v}$ is denoted by $\Delta \boldsymbol{v}=\tilde{\boldsymbol{v}}-\boldsymbol{v}$. The measurement error vector caused by the deviated velocity is denoted by $\tilde{\boldsymbol{r}}$ and is
\begin{align}\label{eq:resdV}
\tilde{\boldsymbol{r}}
=\left[
\begin{matrix}
\boldsymbol{0}_M\\
\left\Vert \boldsymbol{p}_1 - \boldsymbol{p} - \tilde{\boldsymbol{v}}  \delta t_1\right\Vert  -\left\Vert \boldsymbol{p}_1 - {\boldsymbol{p}} - {\boldsymbol{v}}  \delta t_1\right\Vert \\
\vdots\\
\left\Vert \boldsymbol{p}_M - \boldsymbol{p} - \tilde{\boldsymbol{v}}  \delta t_M\right\Vert  -\left\Vert \boldsymbol{p}_M - {\boldsymbol{p}} - {\boldsymbol{v}}  \delta t_M\right\Vert
\end{matrix}
\right]
\text{,}
\end{align}

We denote the estimation bias by $\tilde{\boldsymbol{\mu}}$ and have
\begin{equation}\label{eq:dPvsdV}
\tilde{\boldsymbol{\mu}}=(\tilde{\bm{G}}^T\bm{W}\tilde{\bm{G}})^{-1}\tilde{\bm{G}}^T\bm{W}\tilde{\boldsymbol{r}} \text{,}
\end{equation}
where
$\tilde{\bm{G}}=\bm{G}_\text{Mode 1}$.

We then come to
\begin{equation} \label{eq:mudvsquare}
	\Vert\tilde{\boldsymbol{\mu}}\Vert^2=\tilde{\boldsymbol{r}}^T\bm{S}^T\bm{S}\tilde{\boldsymbol{r}} \text{,}
\end{equation}
where $\bm{S}=(\tilde{\bm{G}}^T\bm{W}\tilde{\bm{G}})^{-1}\tilde{\bm{G}}^T\bm{W}$.


The RMSE, denoted by ${\stackon[-8pt]{{$RMSE$}}{\vstretch{1.6}{\hstretch{8.2}{\tilde{\phantom{\;}}}}}}$, is
\begin{equation} \label{eq:RMSEdV}
	{\stackon[-8pt]{{$RMSE$}}{\vstretch{1.6}{\hstretch{8.2}{\tilde{\phantom{\;}}}}}} =\sqrt{\Vert\tilde{\boldsymbol{\mu}}\Vert^2+\mathrm{tr}\left({\bm{Q}}\right)} \text{,}
\end{equation}
where $\bm{Q}$ is the estimation error variance and $\bm{Q}=\left(\bm{G}^T\bm{W}\bm{G}\right)^{-1}$.

The estimated position error, denoted by ${\stackon[-8pt]{{$RMSE$}}{\vstretch{1.6}{\hstretch{8.2}{\tilde{\phantom{\;}}}}}}_p$, and the clock offset error, denoted by ${\stackon[-8pt]{{$RMSE$}}{\vstretch{1.6}{\hstretch{8.2}{\tilde{\phantom{\;}}}}}}_b$, are
\begin{align} \label{eq:perrorvsv}
	{\stackon[-8pt]{{$RMSE$}}{\vstretch{1.6}{\hstretch{8.2}{\tilde{\phantom{\;}}}}}}_p =\sqrt{\left\Vert[\tilde{\boldsymbol{\mu}}]_{1:N}\right\Vert^2+\mathrm{tr}\left([{\bm{Q}}]_{1:N,1:N}\right)}\text{,}
\end{align}
and
\begin{align}\label{eq:berrorvsv}
	{\stackon[-8pt]{{$RMSE$}}{\vstretch{1.6}{\hstretch{8.2}{\tilde{\phantom{\;}}}}}}_b =\sqrt{[\tilde{\boldsymbol{\mu}}]_{N+1}^2+[{\bm{Q}}]_{N+1,N+1}}\text{.}
\end{align}

We note that when $\tilde{\boldsymbol{v}}=\boldsymbol{0}$, Mode 1 reduces to the conventional method that only estimates $\boldsymbol{p}$, $b$, and $\omega$ such as presented in \cite{zheng2010joint,vaghefi2015cooperative,zou2017joint}. The errors of TWLAS Mode 1 given by (\ref{eq:perrorvsv}) and (\ref{eq:berrorvsv}) are the same as that of the CTWLAS when applying to a moving UD. Therefore, the conventional method CTWLAS that only estimates $\boldsymbol{p}$, $b$, and $\omega$ can be considered as a special case of Mode 1 of the TWLAS method when the UD is stationary.


\section{Numerical Evaluation} \label{simulation}
\subsection{Simulation Settings}
We conduct numerical simulations to evaluate the localization and synchronization performance of the proposed TWLAS method. We compute the RMSE of the position and clock offset estimation results as given by
\begin{align} \label{eq:RMSEpos}
	\text{Position RMSE}&=\sqrt{\frac{1}{N_s}\sum_{1}^{N_s}\Vert\boldsymbol{p}-\hat{\boldsymbol{p}}\Vert^2} \text{,}
\end{align}
and
\begin{align} \label{eq:RMSEclockb}
	\text{Clock offset RMSE}&=\sqrt{\frac{1}{N_s}\sum_{1}^{N_s}\left(b-\hat{b}\right)^2} \text{,}
\end{align}
where $N_s$ is the total number of positioning result samples from the simulation, and $\hat{\boldsymbol{p}}$ and $\hat{b}$ are the localization and synchronization results, respectively, given by the proposed algorithm. We use the CRLB as an accuracy metric for comparison.

We create a simulation scene to evaluate the performance of the proposed TWLAS method. Four ANs are placed on the middle of the four sides of a 600 m$\times$600 m square area. The coordinate of the four ANs are AN \#1 (-300, -300) m, AN \#2 (-300, 300) m, AN \#3 (300, 300) m, AN \#4 (-300, 300) m, respectively. The moving UD is randomly placed inside the square area with four vertices of (-250, -250) m, (-250, 250) m, (250, 250) m, and (-250, 250) m.

At each simulation run, the UD transmits the request signal and the ANs receive to form request-TOA measurements. Then the ANs transmit the response signals in a sequential manner. The UD receives the response signal from AN \#1 after 10 ms delay and then with an incremental of 10 ms to receive the response signals from each of the remaining ANs. The initial value of the UD clock offset and drift are set randomly at the start of each simulation run. The clock offset is drawn from the uniform distribution $b\sim \mathcal{U}(-1,1)$ s, which is a large clock offset range for localization and synchronization. The clock drift is selected from $\omega\sim \mathcal{U}(-10,10)$ parts per million (ppm), which is at the level of the frequency stability of a commonly used temperature compensated oscillator (TCXO). The UD velocity at each simulation run is randomly selected, with its norm $\Vert\boldsymbol{v}\Vert$ drawn from $\mathcal{U}(0,50)$ m/s, and the direction angle drawn from $\mathcal{U}(0,2\pi)$. The clock and motion of the UD evolve during each simulated period based on the clock model and motion model given by (\ref{eq:clockbomega}) and (\ref{eq:posvel}), respectively.

For the iterative TWLAS algorithm, we set the maximum iteration count to $iter=10$, and the convergence threshold to $thr=\sigma/10$. In other words, the algorithm will stop if the number of iterations reaches 10 or the norm of the parameter error $\left\Vert[\check{\boldsymbol{\theta}}]_{1:N}\right\Vert$ is smaller than $\sigma/10$.

%


\subsection{Localization and Synchronization Performance}
The TOA measurement noise $\sigma$ and $\sigma_i$ are set identical. They both vary from 0.01 m to 10 m with 6 steps in total. At each step, 40,000 times of Monte-Carlo simulations are run to generate the random UD position and motion, the clock offset and drift, and the TOA measurements. The simulated data are input to the iterative TWLAS algorithm proposed in Section \ref{locmethod}. The initial parameter $\check{\boldsymbol{\theta}}_{0}$ of the iterative algorithm is set to a random point on the circumference of a 50-m radius circle centered at the true position. The weighting matrix $\bm{W}$ is set using the values of the measurement noise variances $\sigma^2$ and $\sigma_i^2$ based on \eqref{eq:matW}.

\begin{figure}
	\centering
	\includegraphics[width=0.99\linewidth]{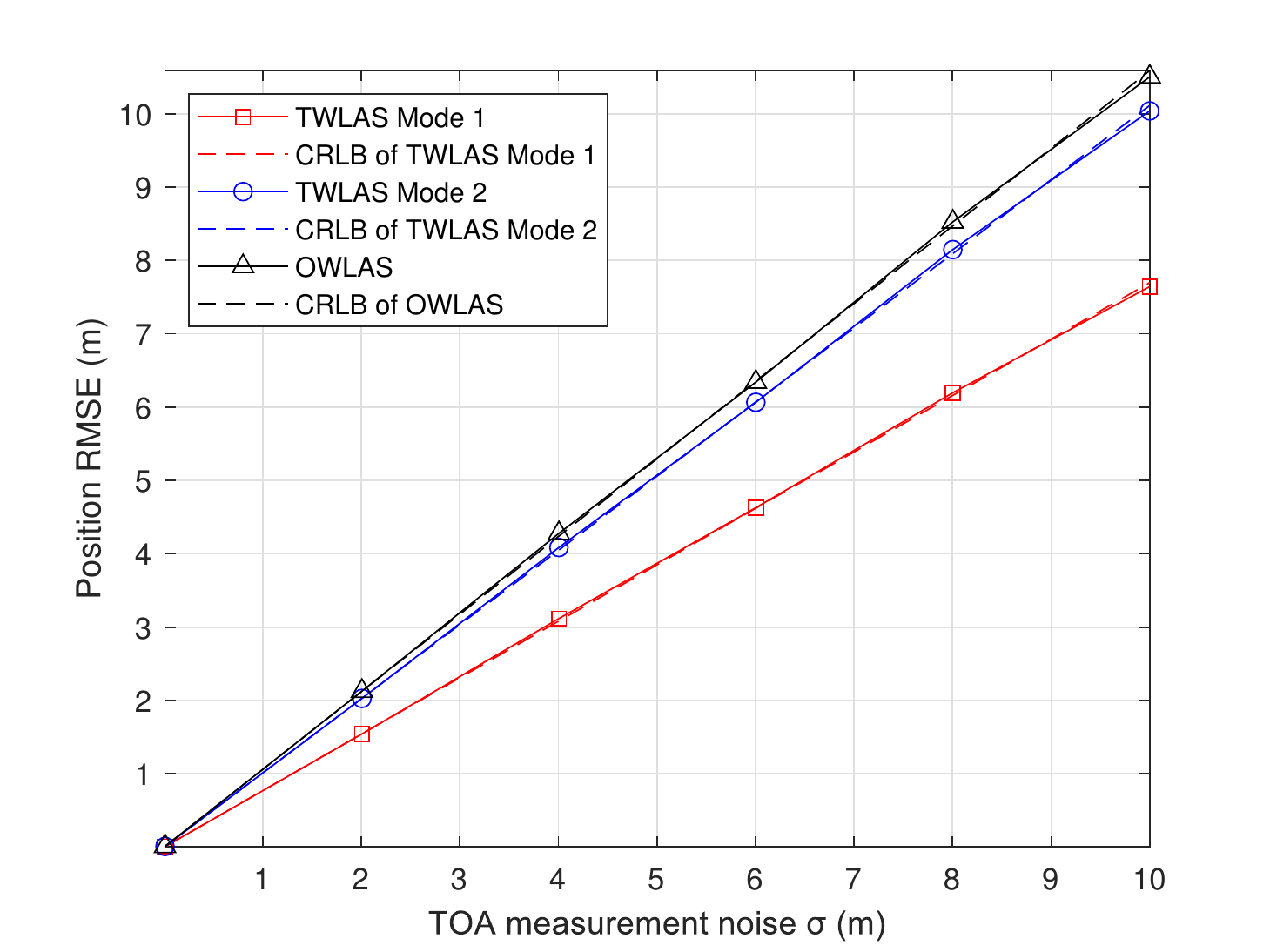} 
	\caption{Localization error vs. measurement noise. The position estimation errors of the two modes of the proposed TWLAS method reach CRLB. The accuracy of Mode 1 is better than that of Mode 2. Both modes of the TWLAS method outperform the conventional one-way TOA method (OWLAS).
	}
	\label{fig:presult}
\end{figure}

\begin{figure}
	\centering
	\includegraphics[width=0.99\linewidth]{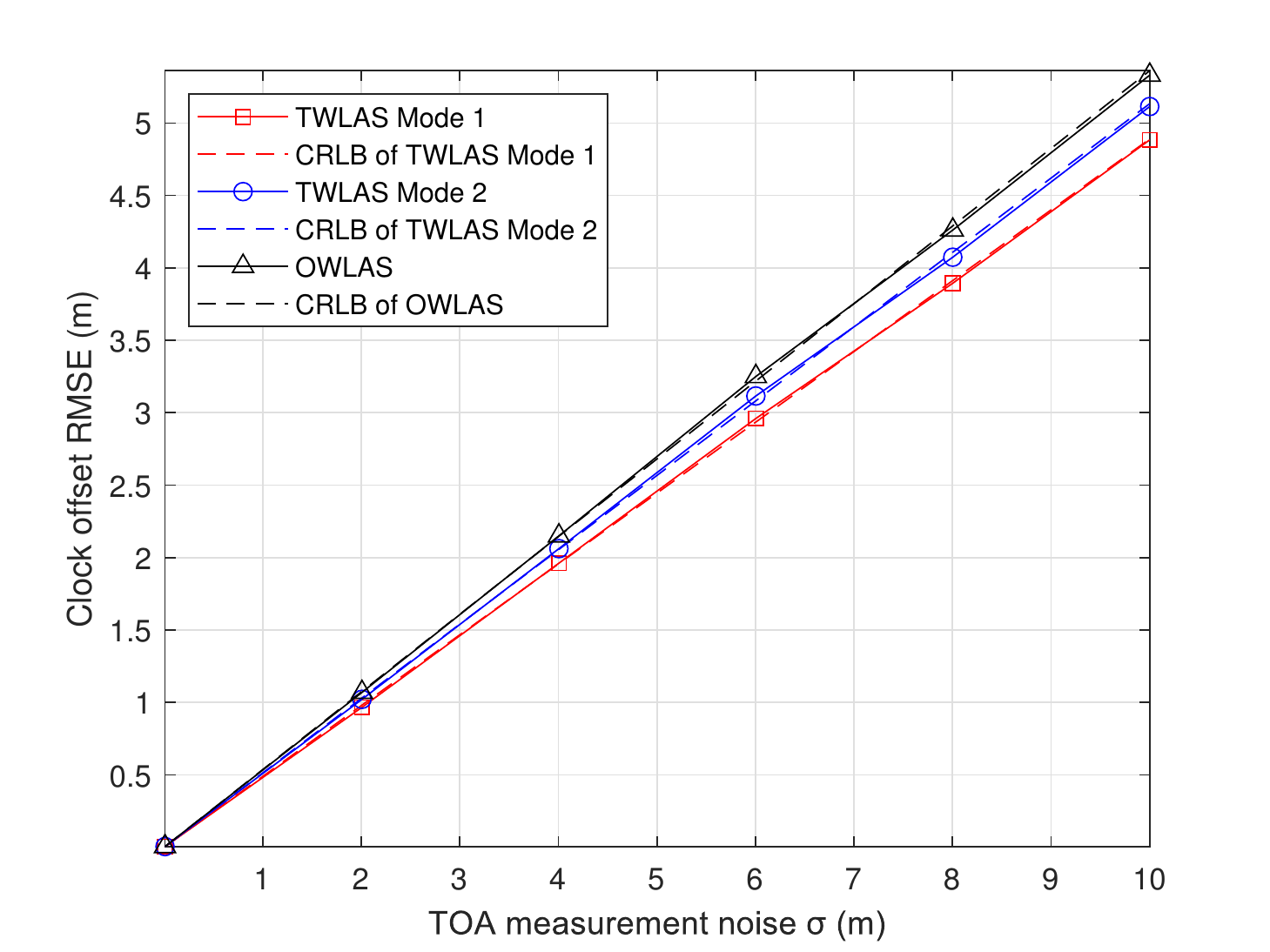} 
	\caption{Synchronization error vs. measurement noise. The clock offset estimation errors of both two modes of the proposed TWLAS method reach their respective CRLBs. The error of Mode 1 is smaller than that of Mode 2. Both modes outperform the conventional one-way TOA method (OWLAS).
	}
	\label{fig:bresult}
\end{figure}

The UD position estimation errors of the two modes of the TWLAS method are shown in Fig. \ref{fig:presult}. Their respective CRLBs are also shown in the same figure. We can see that position errors of both modes reach the theoretical lower bounds, showing their optimality. We also use the conventional one-way TOA localization method (OWLAS) \cite{foy1976position,kaplan2005understanding} to generate the localization error for comparison. We can see that both modes of the TWLAS outperform OWLAS, consistent with the theoretical analysis in Section \ref{locanalysis}. The clock offset estimation errors of the TWLAS are shown in Fig. \ref{fig:bresult}. Similar to Fig. \ref{fig:presult}, both modes of TWLAS outperform OWLAS. The conventional two-way TOA method (CTWLAS) is not compared because it has larger estimation error for a moving UD as will be shown in Section \ref{mode2andold}. The results verify the theoretical analysis presented in Section \ref{locanalysis}.

\begin{figure}
	\centering
	\includegraphics[width=0.99\linewidth]{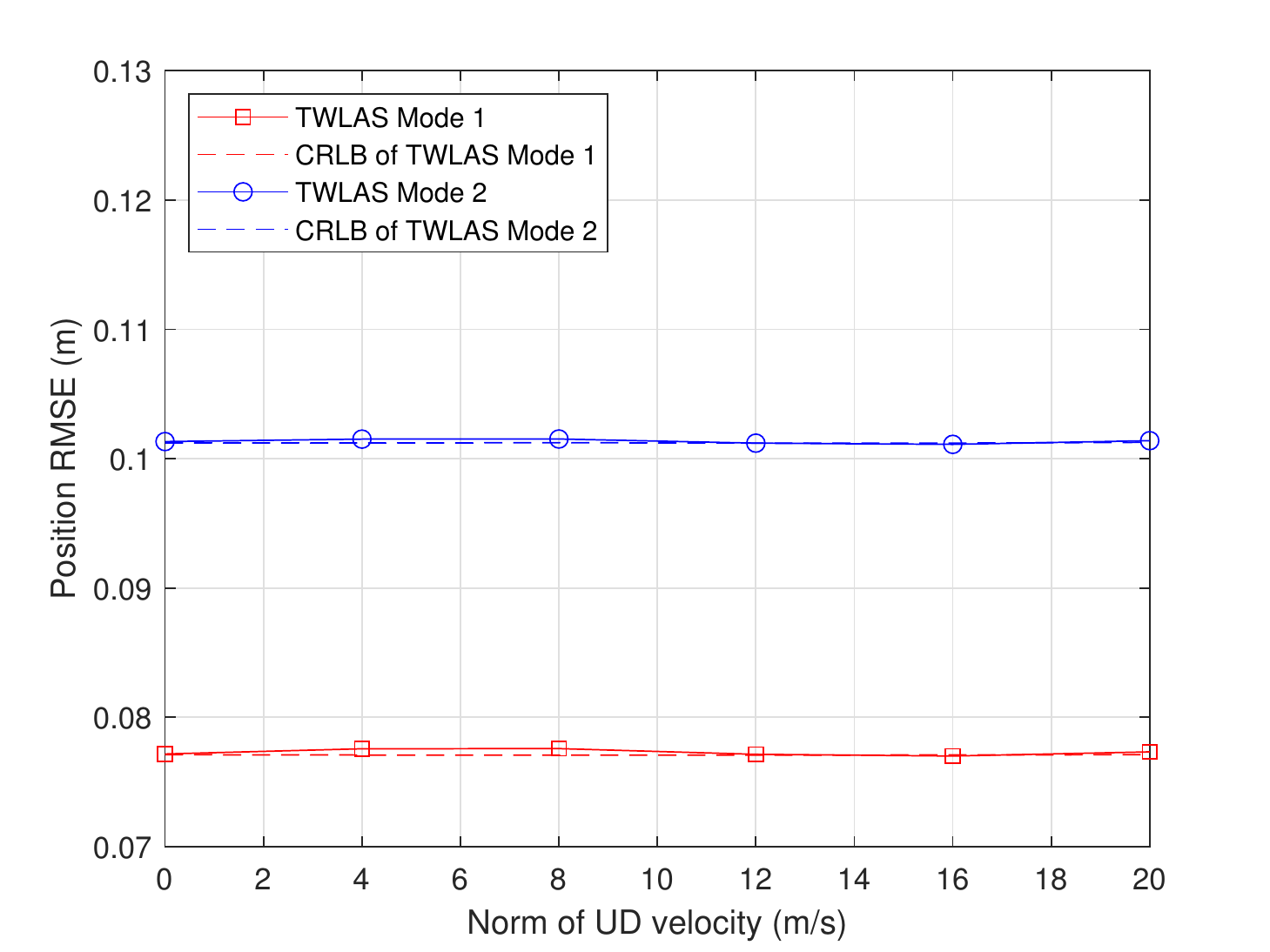} 
	\caption{Localization error vs. norm of UD velocity. The UD velocity information for Mode 1 equals to the true velocity. The measurement noise is $\sigma=0.1$ m. The localization errors of both modes reach CRLB. The accuracy of Mode 1 is better than that of Mode 2 due to the aiding information of the UD velocity. The estimation errors of both modes are irrelevant to the UD motion.
	}
	\label{fig:presultv}
\end{figure}

\begin{figure}
	\centering
	\includegraphics[width=0.99\linewidth]{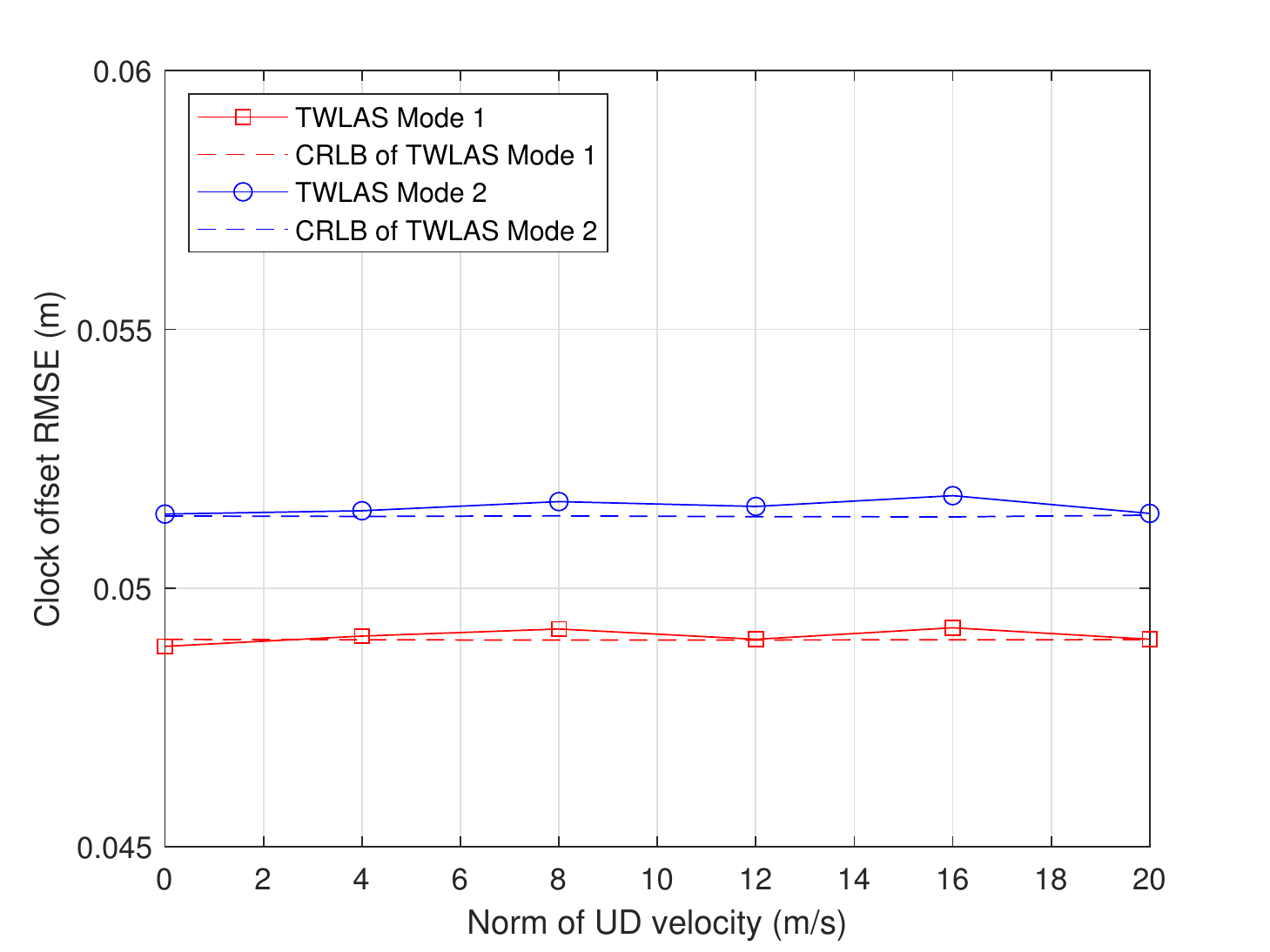} 
	\caption{Synchronization error vs. norm of UD velocity. The UD velocity information for Mode 1 is the true velocity. The measurement noise is $\sigma=0.1$ m. The clock offset estimation errors of both modes reach their respective CRLB. The accuracy of Mode 1 is better than that of Mode 2 due to the aiding information of the UD velocity. The clock offset estimation errors of both modes are irrelevant to the UD motion.
	}
	\label{fig:bresultv}
\end{figure}

We also plot the localization and synchronization error versus the UD speed in Fig. \ref{fig:presultv} and Fig. \ref{fig:bresultv}, respectively, with a fixed measurement noise $\sigma=0.1$ m, which is at the level of a UWB localization device \cite{shi2019blas}. We can see that the estimation errors of both modes are irrelevant to the UD velocity, showing the superiority of the TWLAS method in localization and synchronization for a moving UD. The estimation accuracy of Mode 1 with accurately known UD velocity is better than that of Mode 2, consistent with the error analysis presented in Section \ref{locanalysis}.

\subsection{Comparison between TWLAS Mode 1 and CTWLAS} \label{mode2andold}
We investigate the estimation accuracy of TWLAS Mode 1 in comparison with the conventional two-way TOA method (CTWLAS) that only estimates the UD position and clock parameters, i.e., $\boldsymbol{p}$, $b$, and $\omega$ \cite{zheng2010joint,vaghefi2015cooperative,zou2017joint}. We vary the norm of the UD velocity from 0 m/s to 50 m/s with a step of 10 m/s. We run 40,000 runs of simulations at each step. The direction angle is randomly drawn from $\mathcal{U}(0,2\pi)$. The measurement noise is set to $\sigma =\sigma_i=0.1$ m. The interval between successive TOA measurements is set to $\delta t=$ 5, 10, and 20 ms. Simulations are done for each interval. The simulated TOA measurements are input to both TWLAS Mode 1 and CTWLAS.

\begin{figure}
	\centering
	\includegraphics[width=0.99\linewidth]{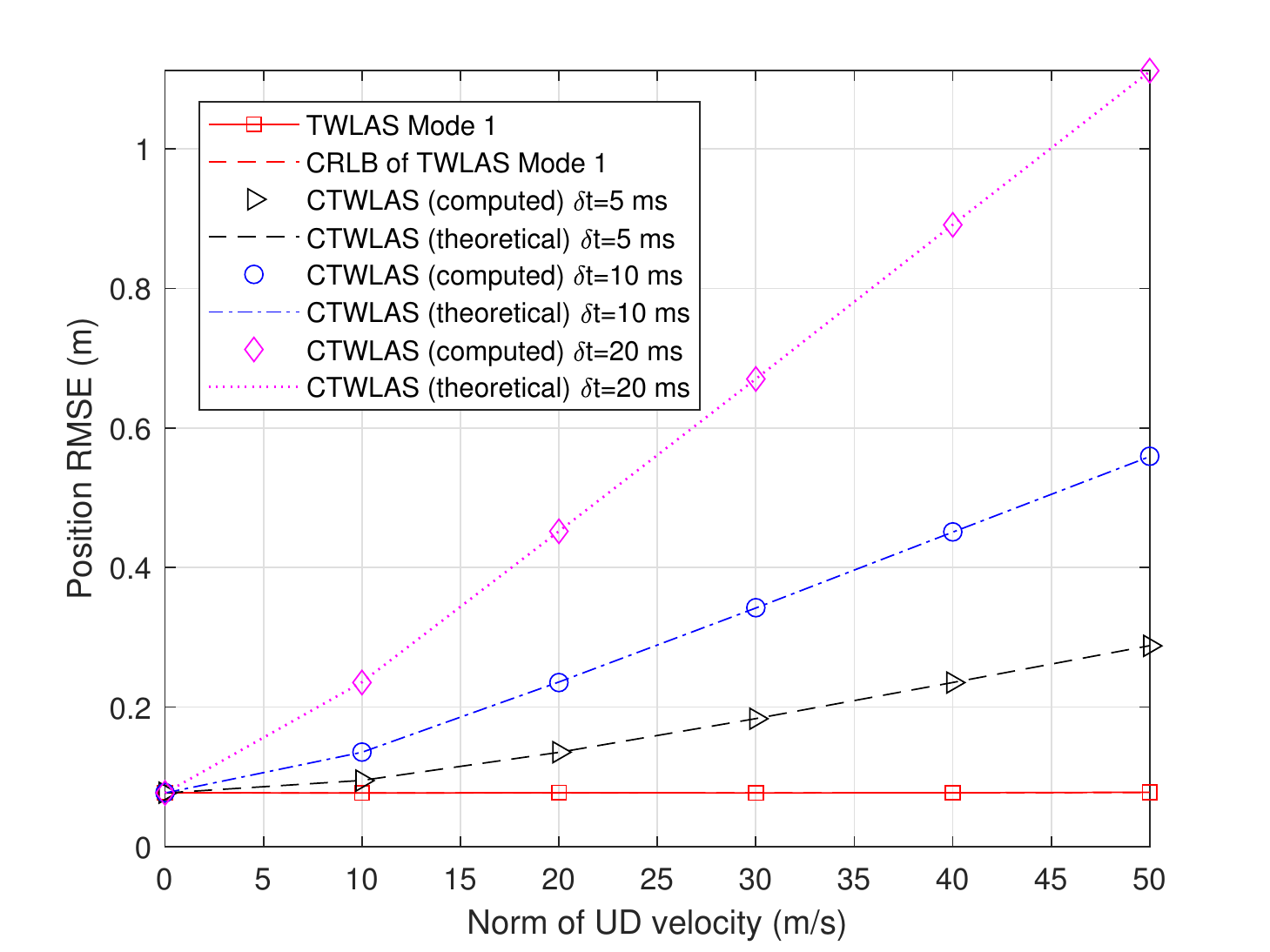} 
	\caption{Localization error comparison between TWLAS Mode 1 and the CTWLAS. The measurement noise is $\sigma=0.1$ m. The localization error of the CTWLAS increases with the growing UD velocity and larger interval $\delta t$, consistent with the theoretical analysis. The error from TWLAS Mode 1 remains stable, showing its superiority.
	}
	\label{fig:presultvold}
\end{figure}

\begin{figure}
	\centering
	\includegraphics[width=0.99\linewidth]{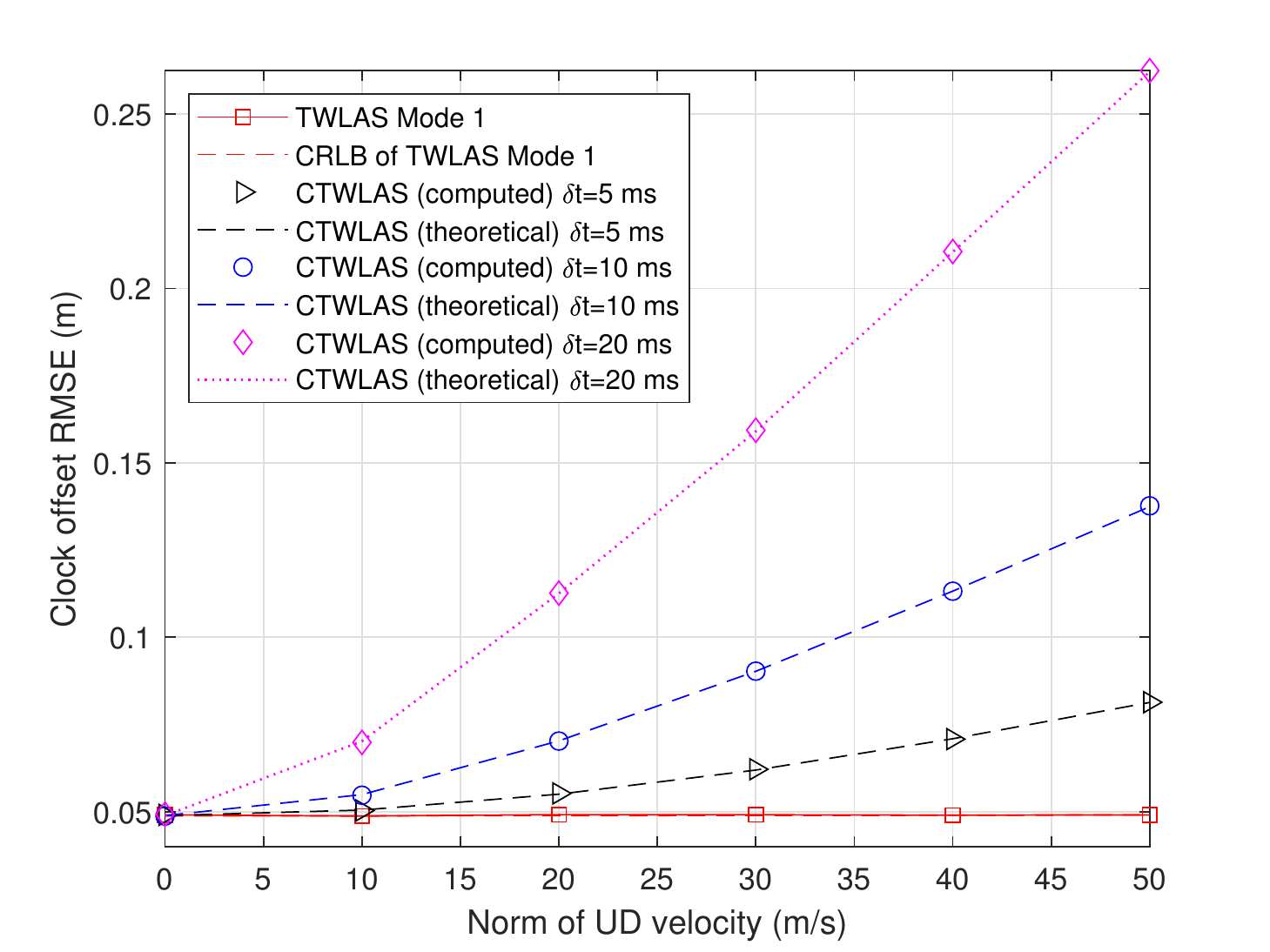} 
	\caption{Synchronization error comparison between TWLAS Mode 1 and the CTWLAS. The measurement noise is $\sigma=0.1$ m. The estimated clock offset error of the CTWLAS increases with the growing UD velocity and larger interval $\delta t$, consistent with the theoretical analysis. TWLAS Mode 1 remains stable when the UD velocity increases, showing its superiority over the CTWLAS.
	}
	\label{fig:bresultvold}
\end{figure}

The localization and synchronization error of both methods are shown in Fig. \ref{fig:presultvold} and Fig. \ref{fig:bresultvold}, respectively. We can see that the estimation error of the CTWLAS grows when the UD velocity increases while that of the proposed TWLAS Mode 1 remains stable regardless of the UD velocity. In addition, the estimation error of the CTWLAS increases with a larger interval $\delta t$. Since the position and clock offset error of the TWLAS remains the same, we only show one curve in each figure. The results are consistent with the theoretical analysis presented in Section \ref{compareold}. It shows the superior performance of the TWLAS over the CTWLAS in localization and synchronization for a moving UD.

\subsection{Performance of TWLAS Mode 1 with Deviated UD Velocity} \label{perfromancedV}
In the case when the known UD velocity deviates from the true value, we evaluate the impact of the deviation on the localization and synchronization error of Mode 1 of the proposed TWLAS method. In the simulation, we vary the norm of the UD velocity deviation $\Vert\Delta \boldsymbol{v}\Vert$ from 0 m/s to 20 m/s with a step of 4 m/s. We run 40,000 simulations at each step. For each simulation run, the direction of the deviated velocity $\Delta \boldsymbol{v}$ is randomly selected from $\mathcal{U}(0,2\pi)$. The velocity input to the TWLAS Mode 1 is set to $\tilde {\boldsymbol{v}}={\boldsymbol{v}}+\Delta\boldsymbol{v}$. The TOA measurement noise $\sigma$ is set to 0.1 m.

\begin{figure}
	\centering
	\includegraphics[width=0.99\linewidth]{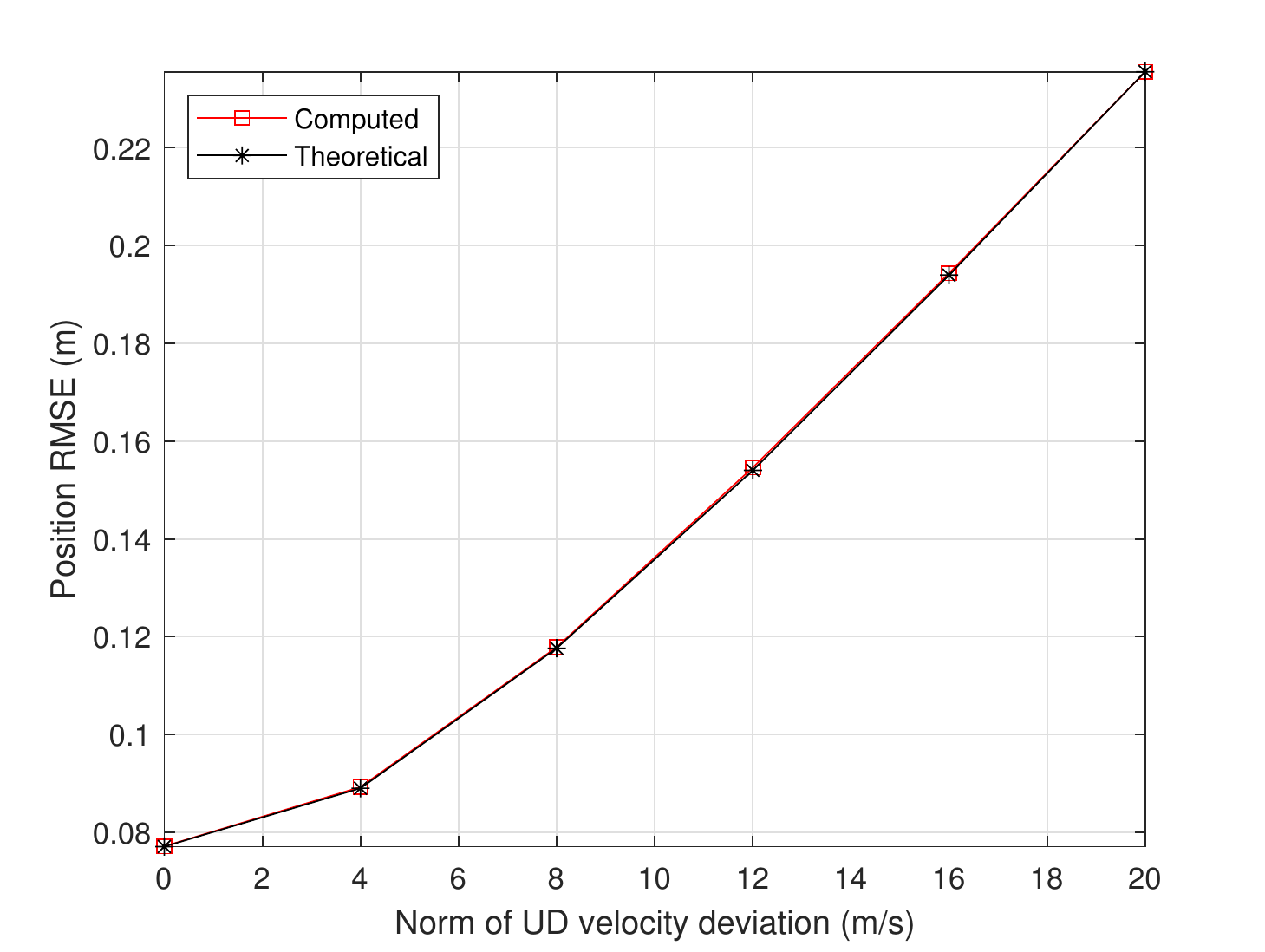} 
	\caption{Estimated position error vs. norm of the UD velocity deviation for Mode 1 of the proposed TWLAS method. The position RMSE from the TWLAS method matches the theoretical analysis presented in Section \ref{locanalysis}.
	}
	\label{fig:perrorvsdVnorm}
\end{figure}

\begin{figure}
	\centering
	\includegraphics[width=0.99\linewidth]{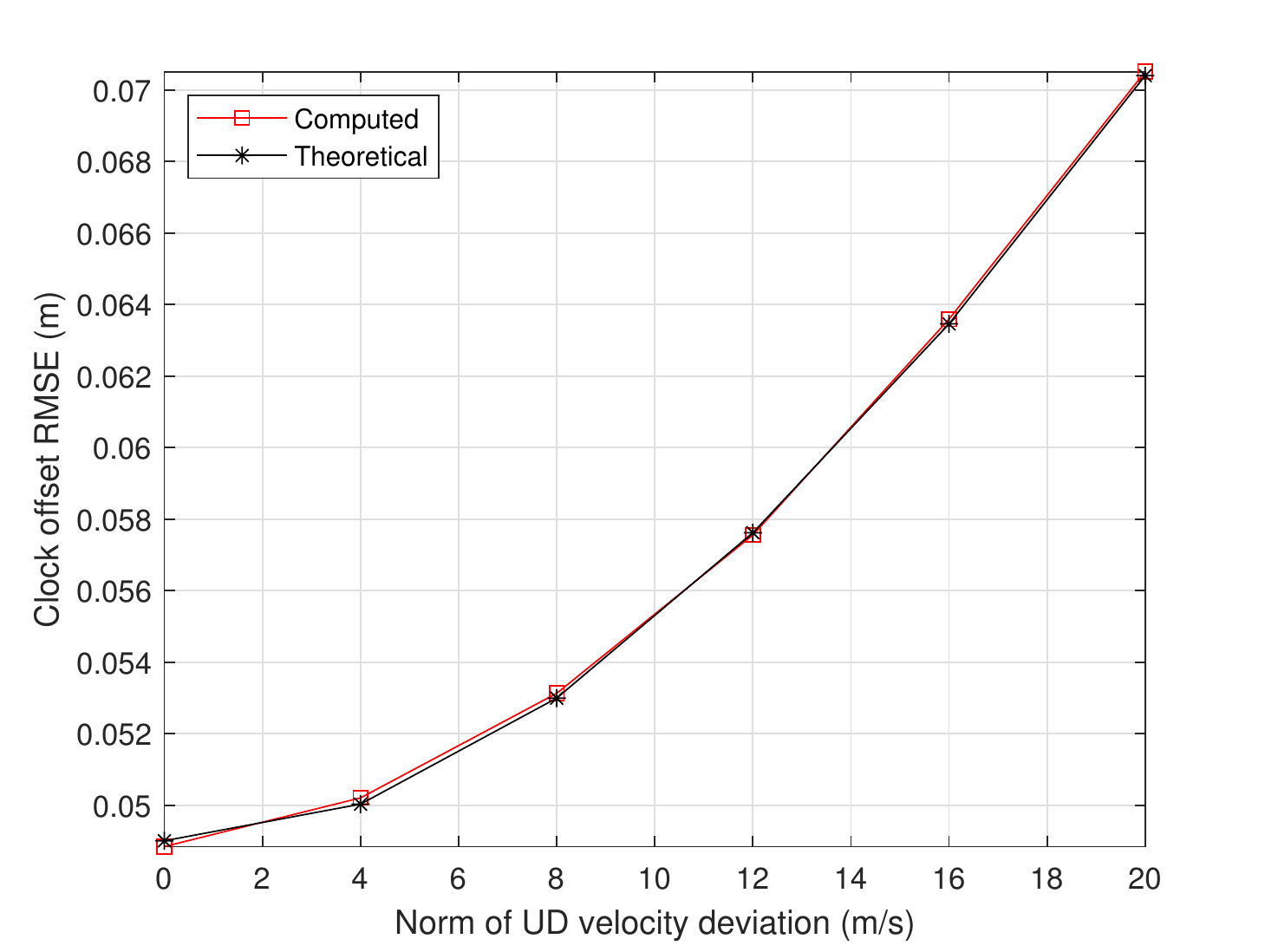} 
	\caption{Estimated clock offset error vs. norm of the UD velocity deviation for Mode 1 of the proposed TWLAS method. The clock offset error estimated by the TWLAS method is consistent with the theoretical analysis.
	}
	\label{fig:berrorvsdV}
\end{figure}

The estimated position and the clock offset error results from Mode 1 of the proposed TWLAS are shown in Fig. \ref{fig:perrorvsdVnorm} and Fig. \ref{fig:berrorvsdV}, respectively. We can see that the estimation errors increase with growing norm of the UD velocity deviation. We also compute the theoretical localization and synchronization errors based on (\ref{eq:perrorvsv}) and (\ref{eq:berrorvsv}), respectively, and plot them in the black star curves in the two figures. We can see that both the estimated position and clock offset RMSEs match the theoretical value. These results verify the theoretical analysis presented in Section \ref{locanalysis}.


The theoretical analysis in Section \ref{locanalysis} can be used as a guidance to evaluate in what cases Mode 1 or Mode 2 can be adopted for UD localization and synchronization in real-world applications. For example, when the UD is moving fast and its velocity information is difficult to obtain, Mode 2 can be used to generate position and clock offset results without being impacted from the UD motion.

\subsection{Dependency of Iterative TWLAS Algorithm on Initial Parameter}
We conduct simulations to evaluate the sensitivity of the proposed iterative algorithm to the initialization error. We set the measurement noise $\sigma=5$ m. In the initial parameter $\check{\boldsymbol{\theta}}_{0}$, we set the initial clock offset $\check{b}_0 = \tau_1$, the initial clock drift $\check{\omega}_0=0$. For Mode 2, we set $\check{\boldsymbol{v}}_0=0$. The initial position $\check{\boldsymbol{p}}_0$ is of most interest. It is randomly selected from the circumference of a circle centered at the true position. The radius of the circle is the initial position error. To evaluate the correctness of the solution from the iterative TWLAS algorithm, we use a threshold of $6\sqrt{\mathsf{CRLB}}$, which equals the 6-$\sigma$ or 99.9999998\% of a Gaussian distribution. In other words, if the estimated position RMSE is smaller than $6\sqrt{\mathsf{CRLB}}$, the result is determined to be correct.

The success rate results under different initial position errors are listed in Table \ref{table_successrate}. As can be seen, with small error in the initial parameter, the iterative TWLAS algorithm is robust and gives 100\% correct results. When the initial position error increases, the algorithm begins to output incorrect solutions with a small probability, especially in Mode 2. We can also observe that Mode 1 is more robust than Mode 2 with an increasing initialization error. The reason is that the known velocity for Mode 1 provides more information and strengthens the problem. In order to obtain the correct and optimal solution, we need an initial parameter as accurate as possible. Some estimation methods such as those based on augmented variables to formulate a closed-form approach \cite{chan1994simple,bancroft1985algebraic}, can be developed to provide a proper initial guess for the iterative algorithm.

\begin{table}[!t]
	\centering
\begin{threeparttable}
	\caption{Success Rate of Iterative TWLAS Algorithm with Initial Parameter Error}
	\label{table_successrate}
	\centering
	\begin{tabular}{c p{1.4cm}  p{1.4cm} p{1.4cm} p{1.4cm} }
		\toprule
		\multirow{2}{1cm}{TWLAS} 
		&\multicolumn{4}{c}{Initial position error (m)}\\
		\cline{2-5}
		\multirow{2}{*}{} & 10& 50& 100&200\\
		
		\hline
		Mode 1  &100\%& 100\%& 100\%& 99.98\%\\
		Mode 2  &100\%& 100\%& 99.98\%& 98.83\%\\
		\bottomrule
	\end{tabular}
	
	\begin{tablenotes}[para,flushleft]
	Note: The TOA measurement noise is set to $\sigma=5$ m. The initial position error is the distance from the true position. For each initial position error, 100,000 simulation runs are done. The localization success rates of both TWLAS Mode 1 and Mode 2 are 100\% when the initial position is not too far from the true position. With increasing initialization error, both modes have decreasing success rates. Mode 1 is less sensitive to the increasing initial error since it has more available information (known UD velocity) than Mode 2.
	\end{tablenotes}
\end{threeparttable}
\end{table}

\subsection{Computational Complexity}
It can be seen from the TWLAS algorithm as given by Algorithm 1 that each iteration has the same operations and thus has the same computational complexity. For each iteration, the major operations are the matrix multiplication and inverse in \eqref{eq:leastsquare}. The complexity of these operations is on the order of $K^3$ \cite{quintana2001note}, where $K$ is the dimension of the matrix. If there are $L$ iterations, the total complexity is on the order of $LK^3$. Note from \eqref{eq:matG} that the design matrix of Mode 1 has fewer columns than Mode 2. Therefore, the complexity of Mode 1 is lower than that of Mode 2.

\begin{figure}
	\centering
	\includegraphics[width=0.99\linewidth]{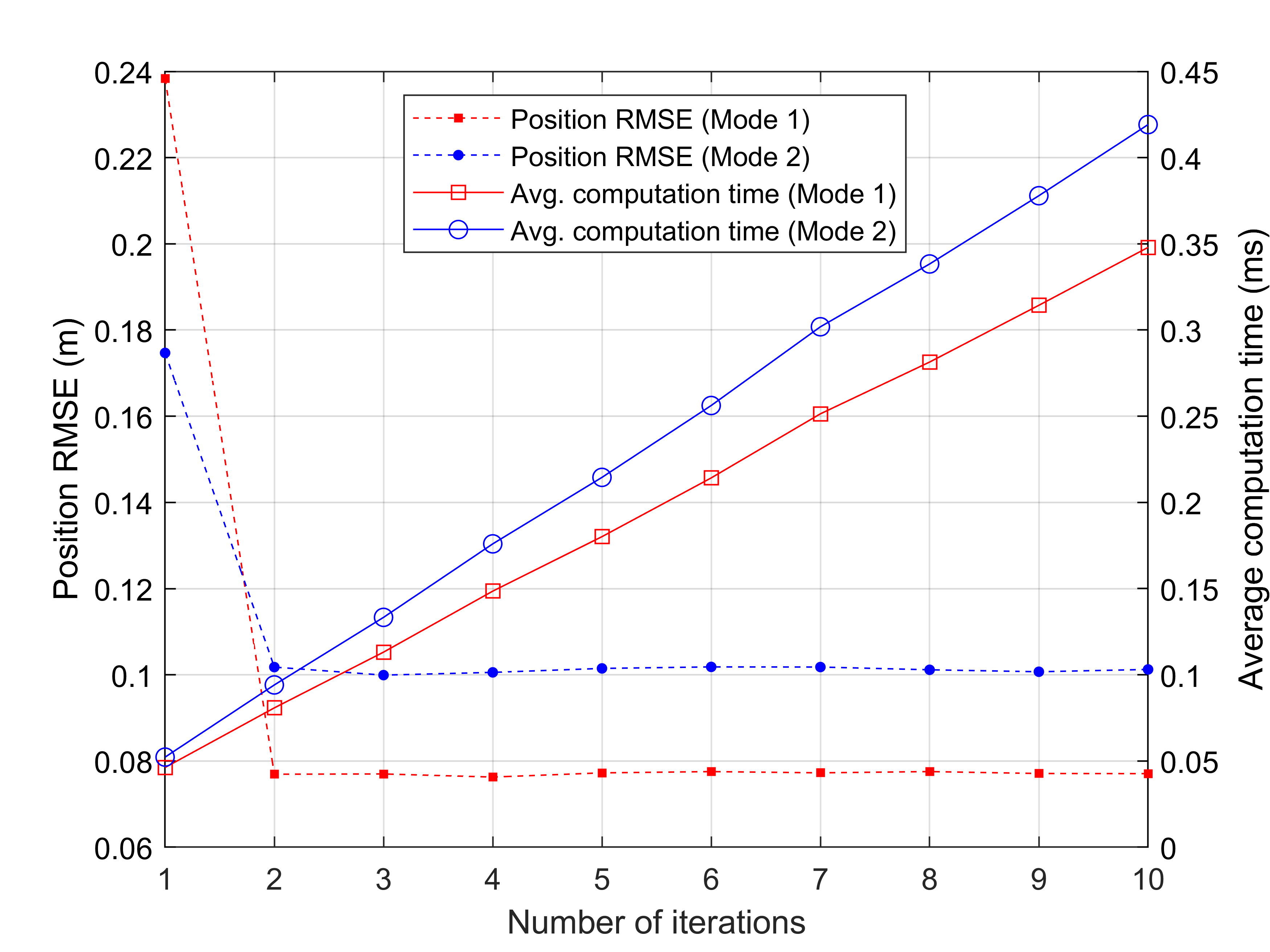} 
	\caption{Position RMSE (dashed lines) and computation time (solid lines) vs. number of iterations. For each number of iterations, the computation time is the average of 10,000 simulation runs. The computation time grows linearly with increasing number of iterations. The position RMSE decreases with an increasing number of iterations and becomes stable quickly after 2nd iteration.
	}
	\label{fig:computationtime}
\end{figure}

We conduct numerical simulations to investigate how the computational complexity and position RMSE change with the number of iterations. The computational platform to run the simulations is a PC with Intel Core i5-10600K CPU @ 4.10 GHz and 32-GB RAM. We set the TOA measurement noise $\sigma=$0.1 m, and the initial position error to 50 m. We vary the number of iterations $L$ from 1 to 10. For each $L$, we run 10,000 Monte-Carlo simulations and compute the average sum the computation time of the iterative algorithm as shown in Fig. \ref{fig:computationtime}. The position RMSE is shown in the same figure. We can see that the computation time has a linear relation with the growing number of iterations. We can see that the computation time has a linear relation with the growing number of iterations. Even with up to 10 iterations, the computation time of the algorithm is only about 0.42 ms for TWLAS Mode 2. Such a low complexity shows that the iterative TWLAS algorithm is applicable on consumer-level electronics platforms such as IoT devices, wearables, drones and robotics.

The position estimation RMSE decreases with an increasing number of iterations and remains stable after 2 iterations as shown in Fig. \ref{fig:computationtime}. In other words, more iterations only cost more computation time, but hardly brings accuracy improvement. For this reason, we expect to exit the algorithm early if the position RMSE becomes stable.

\begin{figure}
	\centering
	\includegraphics[width=0.99\linewidth]{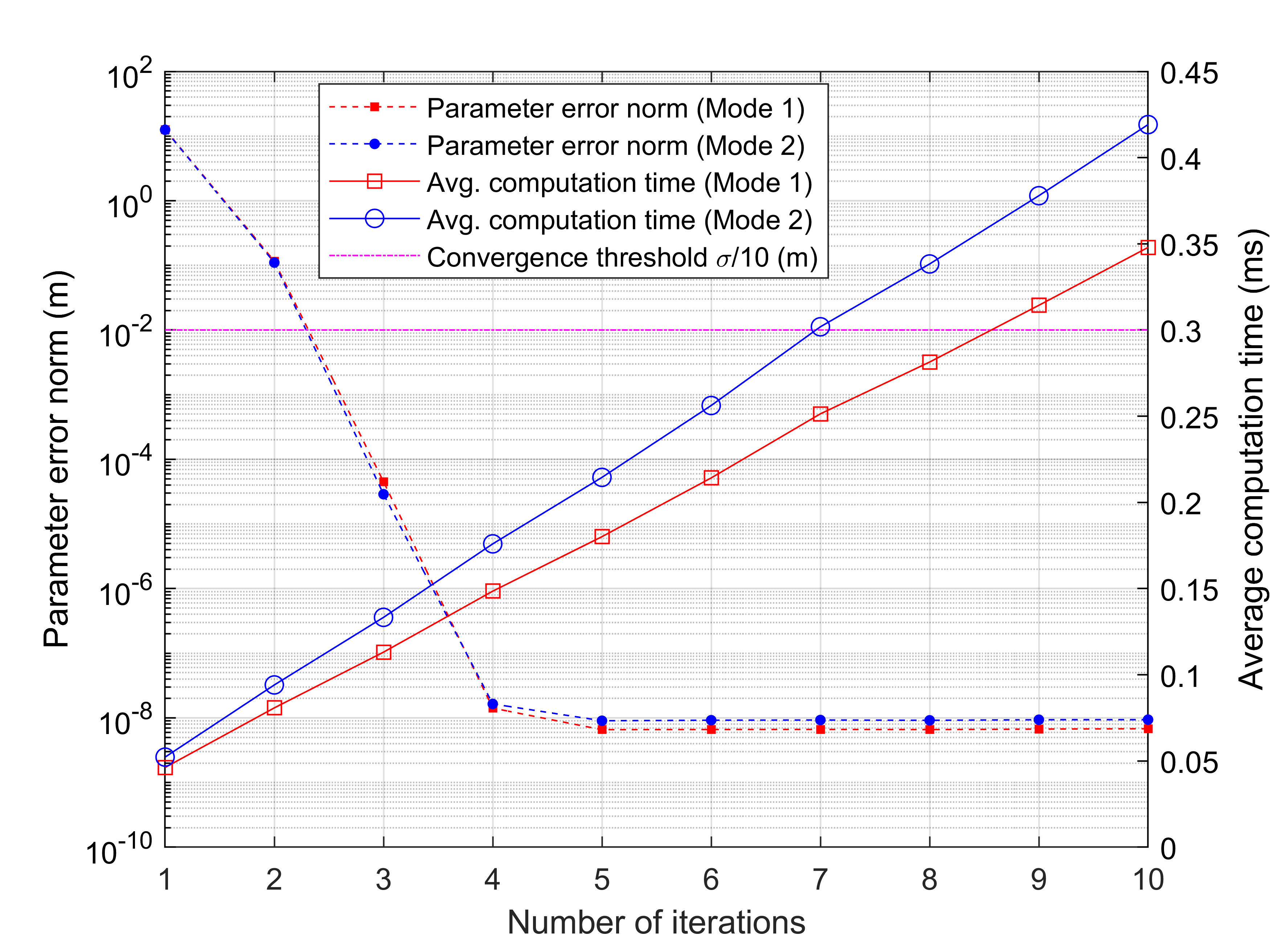} 
	\caption{Parameter error norm (dashed lines) and computation time (solid lines) vs. number of iterations. The convergence threshold (dash-dot line) is attached to the left $y$-axis. The parameter error $\left\Vert [\Delta \check{\boldsymbol{\theta}}]_{1:N} \right\Vert$ decreases with growing number of iterations and becomes smaller than the threshold after 2nd iteration. It becomes a stable small value afterwards.
	}
	\label{fig:computationtimenormx}
\end{figure}

In practice, we do not know the actual position RMSE during the operation of the algorithm. Therefore, we need another proxy variable to indicate whether to continue the iteration or not. As shown in Algorithm 1, we compare the parameter error norm $\left\Vert [\Delta \check{\boldsymbol{\theta}}]_{1:N} \right\Vert$ with $thr$ to decide whether to terminate the algorithm. Using this threshold based criterion, we can reduce the number of iterations and thus reduce complexity. We plot the parameter error in Fig. \ref{fig:computationtimenormx}. It shows that this variable has the same varying trend as the position RMSE shown in Fig. \ref{fig:computationtime}. After the 2nd iteration, it becomes smaller than the threshold $thr$, leading to the early stop of the algorithm. At the same time, as shown in Fig. \ref{fig:computationtime}, the RMSE is stable. Therefore, this result validates that the parameter error is suitable to be an indicator for algorithm termination.


\section{Conclusion}
In this article, we propose an optimal two-way TOA localization and synchronization method, namely TWLAS. Different from existing two-way TOA methods, the new method takes the UD motion into account to compensate the error caused by the UD movement. We analyze its localization and synchronization error and derive the CRLB. The analysis shows that the TWLAS can reach CRLB and has better localization and synchronization accuracy than that of the conventional one-way TOA method. Then conventional two-way TOA method is a special case of the proposed TWLAS method when the UD is stationary. We also derive the relation between the estimation error and the deviated UD velocity information. We conduct Monte-Carlo simulations to evaluate the performance of the proposed TWLAS method. Results show that for a moving UD, the localization and synchronization error of the TWLAS reaches CRLB provided a proper parameter initialization. The accuracy is better than that of the conventional one-way TOA method. The estimation error caused by the deviated UD velocity information is consistent with theoretical analysis.

\appendices

\section{Proof of Theorem \ref{theorem0}}
\label{Appendix1}
The FIMs of Mode 1 and Mode 2 of the TWLAS are
\begin{align}
\mathcal{F}_\text{Mode 1}&=\bm{G}_\text{Mode 1}^T\bm{W}\bm{G}_\text{Mode 1}\nonumber\\
&=\left[
\begin{matrix}
\bm{G}_0^T\bm{W}_{\rho}\bm{G}_0+\bm{G}_1^T\bm{W}_{\tau}\bm{G}_1 & \bm{G}_1^T\bm{W}_{\tau}\boldsymbol{\lambda}\\
\boldsymbol{\lambda}^T\bm{W}_{\tau}\bm{G}_1 & \boldsymbol{\lambda}^T\bm{W}_{\tau}\boldsymbol{\lambda}
\end{matrix}
\right]\text{,}
\end{align}
and
\begin{align}
	\mathcal{F}_\text{Mode 2}&=\bm{G}_\text{Mode 2}^T\bm{W}\bm{G}_\text{Mode 2}\nonumber\\
	&=\left[
	\begin{matrix}
		\bm{G}_0^T\bm{W}_{\rho}\bm{G}_0+\bm{G}_1^T\bm{W}_{\tau}\bm{G}_1 & \bm{G}_1^T\bm{W}_{\tau}\bm{G}_2\\
		\bm{G}_2^T\bm{W}_{\tau}\bm{G}_1 & \bm{G}_2^T\bm{W}_{\tau}\bm{G}_2
	\end{matrix}
	\right]\text{,}
\end{align}
where
$$\boldsymbol{\lambda}=[\delta t_1,\cdots,\delta t_M]^T \text{.}$$

According to (\ref{eq:matG}), we partition $\bm{G}_2$ column-wisely into one column vector and one sub-matrix as $\bm{G}_2$=$\left[ \boldsymbol{\lambda},\bm{L}\right]$, where
$$\bm{L}=\left[
 -\boldsymbol{l}_1\delta t_1,\cdots,
-\boldsymbol{l}_M\delta t_M
\right]^T \text{.}
$$
Then $\mathcal{F}_\text{Mode 2}$ is further derived as
\begin{align}
&\mathcal{F}_\text{Mode 2} \nonumber\\
&=\left[
\begin{matrix}
\bm{G}_0^T\bm{W}_{\rho}\bm{G}_0+\bm{G}_1^T\bm{W}_{\tau}\bm{G}_1 & \bm{G}_1^T\bm{W}_{\tau}\boldsymbol{\lambda} &\bm{G}_1^T\bm{W}_{\tau}\bm{L}\\
\boldsymbol{\lambda}^T\bm{W}_{\tau}\bm{G}_1 & \boldsymbol{\lambda}^T\bm{W}_{\tau}\boldsymbol{\lambda} &\boldsymbol{\lambda}^T\bm{W}_{\tau}\bm{L} \\
\bm{L}^T\bm{W}_{\tau}\bm{G}_1& \bm{L}^T\bm{W}_{\tau}\boldsymbol{\lambda} &\bm{L}^T\bm{W}_{\tau}\bm{L}
\end{matrix}
\right]\nonumber\\
&=\left[\begin{matrix}
\mathcal{F}_\text{Mode 1} &
\begin{matrix}
\bm{G}_1^T\bm{W}_{\tau}\bm{L}\\
\boldsymbol{\lambda}^T\bm{W}_{\tau}\bm{L}
\end{matrix}\\
\begin{matrix}
\bm{L}^T\bm{W}_{\tau}\bm{G}_1& \bm{L}^T\bm{W}_{\tau}\boldsymbol{\lambda}
\end{matrix} & \bm{L}^T\bm{W}_{\tau}\bm{L}
\end{matrix}
\right]
\text{.}
\end{align}

We investigate the top-left $(N+2)\times(N+2)$ sub-matrix in the inverse of $\mathcal{F}_\text{Mode 2}$, which, according to the inverse of a partitioned matrix in \cite{horn2012matrix}, is given by
\begin{align} \label{eq:submatinvF1}
&[\mathcal{F}_\text{Mode 2}^{-1}]_{1:(N+2),1:(N+2)} \nonumber\\
&=\left(\mathcal{F}_\text{Mode 1}-\bm{B}\left(\bm{L}^T\bm{W}_{\tau}\bm{L}\right)^{-1}\bm{B}^T\right)^{-1} \text{,}
\end{align}
where
$$
\bm{B}=\left[
\begin{matrix}
	\bm{G}_1^T\bm{W}_{\tau}\bm{L}\\
	\boldsymbol{\lambda}^T\bm{W}_{\tau}\bm{L}
\end{matrix}
\right] \text{.}
$$

We note that $\bm{A}^T\bm{A}$ is a positive-definite matrix for an arbitrary real matrix $\bm{A}$ with full rank. The matrices $\bm{G}_{\text{Mode 1}}$ and $\bm{G}_{\text{Mode 2}}$ usually have full rank with sufficient ANs that are properly placed. Thus, $\mathcal{F}_\text{Mode 1}$, $\mathcal{F}_\text{Mode 2}$ and $\bm{L}^T\bm{W}_{\tau}\bm{L}$ are all positive-definite. $\bm{B}\left(\bm{L}^T\bm{W}_{\tau}\bm{L}\right)^{-1}\bm{B}^T$ is thereby positive-definite.

Note that $\bm{A}\succ\bm{B}$ if and only if $(\bm{A}-\bm{B})$ is positive-definite. We have
\begin{align} \label{eq:matgreater}
\mathcal{F}_{\text{Mode 1}}\succ\mathcal{F}_{\text{Mode 1}}-\bm{B}\left(\bm{L}^T\bm{W}_{\tau}\bm{L}\right)^{-1}\bm{B}^T \text{.}
\end{align}

We apply inverse to the matrices on both sides of (\ref{eq:matgreater}) and come to
\begin{align} \label{eq:matsmaller}
\mathcal{F}_{\text{Mode 1}}^{-1}\prec\left(\mathcal{F}_\text{Mode 1}-\bm{B}\left(\bm{L}^T\bm{W}_{\tau}\bm{L}\right)^{-1}\bm{B}^T\right)^{-1}.
\end{align}

We are interested in the position and clock offset related terms in the top-left $(N+1)\times(N+1)$ sub-matrix, which is the $(N+1)$-th leading principal minor. We note that all leading principal minors of a positive-definite matrix are positive definite. Thus, we have
\begin{align} \label{eq:submatsmaller}
	\left[\mathcal{F}_{\text{Mode 1}}^{-1}\right]_{1:(N+1),1:(N+1)}\prec\left[\mathcal{F}_{\text{Mode 2}}^{-1}\right]_{1:(N+1),1:(N+1)}.
\end{align}
Thus, for all the diagonal entries of the two matrices, we have
\begin{equation}
	\left[\mathcal{F}_{\text{Mode 1}}^{-1}\right]_{i,i}<\left[\mathcal{F}_{\text{Mode 2}}^{-1}\right]_{i,i}, i=1,\cdots,N+1 \text{,}
\end{equation}
and finish the proof.

\section{Proof of Theorem \ref{theorem1}}
\label{Appendix2}
We derive the top-left $(N+1)\times(N+1)$ sub-matrix of the inverse of $\mathcal{F}_{\text{Mode 2}}$ as
\begin{align} \label{eq:invFmode2}
\left[\mathcal{F}_{\text{Mode 2}}^{-1}\right]_{1:(N+1),1:(N+1)}&=\left(\bm{G}_0^T\bm{W}_{\rho}\bm{G}_0 + \bm{D}\right)^{-1} \nonumber\\
&=\left(\mathcal{F}_{\text{OWLAS}} + \bm{D}\right)^{-1}
\text{,}
\end{align}
where
\begin{align} \label{eq:Dexpression}
\bm{D}=\bm{G}_1^T\bm{W}_{\tau}\bm{G}_1-\bm{G}_1^{T}\bm{W}_{\tau}\bm{G}_2\left(\bm{G}_2^{T}\bm{W}_{\tau}\bm{G}_2\right)^{-1}\bm{G}_2^{T}\bm{W}_{\tau}\bm{G}_1
\text{.}
\end{align}

A special case is that the UD receives the response signals from all the ANs simultaneously, i.e., $\delta t_1=\delta t_2=\cdots=\delta t_M=\delta t$. In this case, the matrix $\bm{G}_2$ becomes
$$
\bm{G}_2=\delta t\left[
\begin{matrix}
	1  & -\boldsymbol{l}_1^T\\
	\vdots & \vdots \\
	1 & -\boldsymbol{l}_M^T
\end{matrix}
\right]=\delta t \bm{G}_1 \bm{P} \text{,}
$$
where
$$
\bm{P}=\left[
\begin{matrix}
	\boldsymbol{0}_{N} &\bm{I}_N \\
	1 &\boldsymbol{0}_{N}^T
\end{matrix}
\right] \text{.}
$$
We plug this expression of $\bm{G}_2$ into (\ref{eq:Dexpression}), and obtain 
\begin{align} \label{eq:Dmatrix}
\bm{D}=&\bm{G}_1^T\bm{W}_{\tau}\bm{G}_1- \nonumber\\
&\bm{G}_1^{T}\bm{W}_{\tau}\delta t\bm{G}_1\bm{P}\left(\delta t\bm{P}\bm{G}_1^{T}\bm{W}_{\tau}\bm{G}_1\bm{P}\delta t\right)^{-1}\delta  t\bm{P}\bm{G}_1^{T}\bm{W}_{\tau}\bm{G}_1 \nonumber\\
=&\bm{G}_1^T\bm{W}_{\tau}\bm{G}_1- \nonumber\\
&\bm{G}_1^{T}\bm{W}_{\tau}\bm{G}_1\bm{P}\bm{P}^{-1}\left(\bm{G}_1^{T}\bm{W}_{\tau}\bm{G}_1\right)^{-1}\bm{P}^{-1}\bm{P}\bm{G}_1^{T}\bm{W}_{\tau}\bm{G}_1 \nonumber\\
=&\bm{G}_1^T\bm{W}_{\tau}\bm{G}_1- \bm{G}_1^{T}\bm{W}_{\tau}\bm{G}_1\left(\bm{G}_1^{T}\bm{W}_{\tau}\bm{G}_1\right)^{-1}\bm{G}_1^{T}\bm{W}_{\tau}\bm{G}_1 \nonumber\\
=&\bm{G}_1^T\bm{W}_{\tau}\bm{G}_1- \bm{G}_1^{T}\bm{W}_{\tau}\bm{G}_1 \nonumber\\
=&\bm{O}_{(N+1)\times(N+1)}\text{,}
\end{align}
where $\bm{O}_{(N+1)\times(N+1)}$ is an all-zero matrix.

We substitute (\ref{eq:Dmatrix}) into (\ref{eq:invFmode2}) and thus have
$$\left[\mathcal{F}_{\text{Mode 2}}^{-1}\right]_{1:(N+1),1:(N+1)}=\mathcal{F}_{\text{OWLAS}}^{-1} \text{,}$$
i.e., TWLAS Mode 2 and OWLAS have identical estimation performance in such a special case with simultaneous response signals from all the ANs.

We then investigate the general case. We note that all the diagonal entries of $\bm{W}_{\tau}$ are equal. Therefore,
\begin{align} \label{eq:D1}
\bm{D}&=\bm{G}_1^T\bm{W}_{\tau}\left(\bm{W}_{\tau}^{-1}-\bm{G}_2\left(\bm{G}_2^{T}\bm{W}_{\tau}\bm{G}_2\right)^{-1}\bm{G}_2^{T}\right)\bm{W}_{\tau}\bm{G}_1\nonumber\\
&=\frac{1}{\sigma^2}\bm{G}_1^T\left(\bm{I}_M-\bm{G}_2\left(\bm{G}_2^{T}\bm{G}_2\right)^{-1}\bm{G}_2^{T}\right)\bm{G}_1
\text{.}
\end{align}

We conduct singular value decomposition on $\bm{G}_2$ and
$$\bm{G}_2=\bm{U}\bm{\Sigma}\bm{V}^T \text{,}$$
where $\bm{U}$ is a $M\times M$ orthogonal matrix, $\bm{\Sigma}$ is an $M\times (N+1)$ diagonal matrix with non-negative diagonal entries, and $\bm{V}$ is a $(N+1)\times (N+1)$ orthogonal matrix.

The matrix in the parenthesis in (\ref{eq:D1}) becomes
\begin{align}
&\bm{I}_M-\bm{G}_2\left(\bm{G}_2^{T}\bm{G}_2\right)^{-1}\bm{G}_2^{T}\nonumber\\
&=\bm{U}\left(\bm{I}_M-\bm{\Sigma}\bm{V}^T\left(\bm{V}\bm{\Sigma}^T\bm{U}^T\bm{U}\bm{\Sigma}\bm{V}^T\right)^{-1}\bm{V}\bm{\Sigma}^T\right)\bm{U}^T\nonumber\\
&=\bm{U}\left(\bm{I}_M-\bm{\Sigma}\bm{V}^T\bm{V}\left(\bm{\Sigma}^T\bm{\Sigma}\right)^{-1}\bm{V}^T\bm{V}\bm{\Sigma}^T\right)\bm{U}^T \nonumber\\
&=\bm{U}\left(\bm{I}_M-\bm{\Sigma}\left(\bm{\Sigma}^T\bm{\Sigma}\right)^{-1}\bm{\Sigma}^T\right)\bm{U}^T 
\end{align}

We note that
\begin{align}
&\bm{\Sigma}\left(\bm{\Sigma}^T\bm{\Sigma}\right)^{-1}\bm{\Sigma}^T=\nonumber\\
&\left[
\begin{matrix}
\bm{I}_{N+1} & \bm{O}_{(N+1)\times (M-N-1)}\\
\bm{O}_{(M-N-1)\times (N+1)} & \bm{O}_{(M-N-1)\times (M-N-1)}
\end{matrix}
\right]\text{.}
\end{align}
Therefore, 
\begin{align}
&\bm{I}_M-\bm{\Sigma}\left(\bm{\Sigma}^T\bm{\Sigma}\right)^{-1}\bm{\Sigma}^T\nonumber\\
&=\left[
\begin{matrix}
	\bm{O}_{(N+1)\times (N+1)} & \bm{O}_{(N+1)\times (M-N-1)}\\
	\bm{O}_{(M-N-1)\times (N+1)} & \bm{I}_{M-N-1}
\end{matrix}
\right]\text{,}
\end{align}
in which all the eigenvalues are non-negative, indicating its positive semi-definiteness. Then $\bm{D}$ is positive semi-definite, i.e., $\bm{D}\succeq 0$. Thus, we come to 
\begin{align} \label{eq:plusD}
\mathcal{F}_{\text{OWLAS}} + \bm{D} \succeq \mathcal{F}_{\text{OWLAS}}
	\text{,}
\end{align}

We take inverse on (\ref{eq:plusD}) and the left side becomes (\ref{eq:invFmode2}). Then, we have
\begin{align}
	\left[\mathcal{F}_{\text{Mode 2}}^{-1}\right]_{1:(N+1),1:(N+1)}\preceq\mathcal{F}_{\text{OWLAS}}^{-1}
	\text{.}
\end{align}
The diagonal elements, which represent the localization and synchronization accuracy, have the relation as given by
\begin{equation}
		\left[\mathcal{F}_{\text{Mode 2}}^{-1}\right]_{i,i}\leq \left[\mathcal{F}_{\text{OWLAS}}^{-1}\right]_{i,i},i=1,\cdots\,N+1 \text{.}
\end{equation}
Thus, we have proved (\ref{eq:Fisher2andOW}).

\ifCLASSOPTIONcaptionsoff
    \newpage
\fi



\bibliographystyle{IEEEtran}
\bibliography{IEEEabrv,paper}
\end{document}